\documentclass[10pt,a4paper,twocolumn]{article}

\usepackage[
  a4paper,
  top=16mm,bottom=17mm,left=16mm,right=16mm,
  headheight=13pt,headsep=7pt,footskip=12pt
]{geometry}
\usepackage{fontspec}
\usepackage{amsmath}
\usepackage{unicode-math}
\usepackage{flushend}
\usepackage{microtype}
\usepackage{titlesec}
\usepackage{needspace}
\usepackage{enumitem}
\usepackage{indentfirst}
\usepackage{xcolor}
\usepackage[numbers,square,sort&compress]{natbib}
\usepackage{hyperref}
\usepackage{xurl}

\newcommand{\papertitle}{Theoretical progress on regularized lattice Boltzmann collision models on weakly incompressible and compressible thermal flows}
\newcommand{\paperauthors}{Xuhui Li, Xiaowen Shan, Wenyang Duan}
\title{\papertitle}
\author{\paperauthors}
\date{}
\renewenvironment{abstract}
  {\noindent\textbf{\abstractname:}\enspace\ignorespaces}
  {\par}

\hypersetup{
  colorlinks=true,
  allcolors=black,
  pdfborder={0 0 0},
  hypertexnames=false,
  pdftitle={\papertitle},
  pdfauthor={\paperauthors},
  pdfsubject={Review of regularized lattice Boltzmann collision models},
  pdfkeywords={lattice Boltzmann method, regularized collision model, numerical stability, spurious mode}
}
\allowdisplaybreaks[2]
\setlist[enumerate]{leftmargin=1.35em,itemsep=0.15em,topsep=0.25em,parsep=0pt}


\titleformat{\section}
  {\normalfont\large\bfseries}{\thesection}{0.45em}{}
\titlespacing*{\section}{0pt}{1.0em plus 0.2em minus 0.1em}{0.35em}
\titleformat{\subsection}
  {\normalfont\normalsize\bfseries}{\thesubsection}{0.4em}{}
\titlespacing*{\subsection}{0pt}{0.65em plus 0.15em minus 0.1em}{0.25em}
\ExplSyntaxOn
\seq_new:N \l__refc_items_seq
\seq_new:N \l__refc_range_seq
\tl_new:N \l__refc_keys_tl
\int_new:N \l__refc_start_int
\int_new:N \l__refc_end_int
\cs_new_protected:Npn \__refc_add_key:n #1
  {
    \tl_if_empty:NF \l__refc_keys_tl { \tl_put_right:Nn \l__refc_keys_tl {,} }
    \tl_put_right:Nn \l__refc_keys_tl {r#1}
  }
\NewDocumentCommand{\refc}{m}
  {
    \tl_clear:N \l__refc_keys_tl
    \seq_set_split:Nnn \l__refc_items_seq {,} {#1}
    \seq_map_inline:Nn \l__refc_items_seq
      {
        \tl_if_in:nnTF {##1} {--}
          {
            \seq_set_split:Nnn \l__refc_range_seq {--} {##1}
            \int_set:Nn \l__refc_start_int {\seq_item:Nn \l__refc_range_seq {1}}
            \int_set:Nn \l__refc_end_int {\seq_item:Nn \l__refc_range_seq {2}}
            \int_step_inline:nnnn
              {\l__refc_start_int} {1} {\l__refc_end_int}
              {\__refc_add_key:n {####1}}
          }
          {\__refc_add_key:n {##1}}
      }
    \unskip\,\exp_args:NV \cite \l__refc_keys_tl
  }
\ExplSyntaxOff
\newcommand{\secref}[1]{Section~\ref{sec:#1}}
\newcommand{\secrefs}[2]{Sections~\ref{sec:#1} and~\ref{sec:#2}}
\newcommand{\et}[1]{\tag{#1}\label{eq:#1}}
\newcommand{\E}[1]{Eq.~\eqref{eq:#1}}
\newcommand{\Es}[2]{Eqs.~\eqref{eq:#1}--\eqref{eq:#2}}
\newcommand{\dd}{\,\mathrm{d}}
\newcommand{\Tr}{\operatorname{Tr}}
\newcommand{\perm}{\operatorname{perm}}
\newcommand{\co}{\symbf{c}}
\newcommand{\xv}{\symbf{\xi}}
\newcommand{\uv}{\symbf{u}}
\newcommand{\gv}{\symbf{g}}
\newcommand{\delv}{\symbf{\delta}}
\newcommand{\Fv}{\symbf{F}}
\newcommand{\Om}{\Omega}
\newcommand{\fbar}{\overline{f}}
\newcommand{\qbar}{\overline{q}}
\newcommand{\abar}{\overline{a}}
\newcommand{\bbar}{\overline{b}}
\newcommand{\dbar}{\overline{d}}

\renewcommand{\boldsymbol}[1]{\symbf{#1}}

\begin{document}

\setcounter{page}{1}

\twocolumn[
\thispagestyle{plain}
\begin{center}
  {\LARGE\normalfont\bfseries\papertitle\par}
  \vspace{0.55em}
  {\large\normalfont Xuhui Li\textsuperscript{1,*}, Xiaowen Shan\textsuperscript{2}, Wenyang Duan\textsuperscript{1}\par}
  \vspace{0.35em}
  {\small\itshape
   \textsuperscript{1}College of Shipbuilding Engineering, Harbin Engineering University, Harbin 150001, China;\par\vspace{-0.15em}
   \textsuperscript{2}Institute for Advanced Study, Beijing Normal-Hong Kong Baptist University, China\par}
\end{center}

\vspace{0.25em}
\begin{minipage}{0.96\textwidth}
\small
\begin{abstract}
Numerical stability has long constrained the application of lattice Boltzmann methods to high-Reynolds-number and high-Mach-number flows. Improving the stability and accuracy of collision models has therefore remained a major challenge for nearly three decades. Regularized collision models constitute an important class of lattice Boltzmann collision models and have recently undergone substantial theoretical development. They are now widely used for high-Reynolds-number turbulence, high-Mach-number compressible flow and aeroacoustics. This article systematically reviews their development and establishes a unified theoretical framework that clarifies the relationships among the principal formulations. The analysis identifies the underlying principle of the regularized multiple-relaxation collision model: higher-order kinetic moments that cannot be represented by the discrete lattice are rapidly relaxed in a coordinate-independent, self-similar moment space, thereby preventing spurious modes from contaminating the hydrodynamic equations.
\end{abstract}

\vspace{0.25em}

\vspace{0.25em}
\par\noindent\textsuperscript{*}Corresponding author: \href{mailto:lixuhui@hrbeu.edu.cn}{lixuhui@hrbeu.edu.cn}.

\end{minipage}
\vspace{0.7em}
]

\setlength{\abovedisplayskip}{5pt plus 2pt minus 5pt}
\setlength{\belowdisplayskip}{6pt plus 2pt minus 2pt}
\setlength{\abovedisplayshortskip}{0pt plus 1pt minus 1pt}
\setlength{\belowdisplayshortskip}{6pt plus 2pt minus 2pt}

\section*{0\quad Introduction}
\phantomsection\label{sec:introduction}
\addcontentsline{toc}{section}{0 Introduction}

Over the past three decades, the lattice Boltzmann method has gradually become an important branch of computational fluid dynamics and has been widely applied to scientific and engineering problems \refc{1--2}. Unlike conventional computational fluid dynamics methods based on discretizing the Navier--Stokes--Fourier (NSF) equations, such as finite-difference and finite-volume methods, the lattice Boltzmann method is a mesoscopic numerical approach that describes a fluid system through the collision and streaming of particle distribution functions. Its connection with the NSF equations can be established through a Chapman--Enskog (CE) expansion \refc{3} in the limit of small Knudsen number ($\mathrm{Kn}$).

For conventional computational fluid dynamics methods, a major difficulty is the treatment of nonlinear convective terms in partial differential equations. High-order discretization schemes are therefore often adopted to control numerical dissipation and dispersion. For the on-lattice velocity sets commonly used by lattice Boltzmann methods, particle populations stream along characteristics and reach neighbouring Cartesian nodes in a single time step. This operation requires no interpolation and amounts computationally to data transfer, so the Lagrangian streaming step simplifies the treatment of convection and introduces little numerical dissipation or dispersion. The lattice Boltzmann method also avoids the iterative solution of a pressure Poisson equation, advances explicitly in time, and evaluates the nonlinear collision term locally in physical space. The resulting computation is predominantly algebraic and has a regular memory-access pattern. These features are well suited to heterogeneous parallel architectures, including graphics processing units (GPUs) and the Sunway many-core architecture, and enable large-scale, high-accuracy fluid simulations \refc{4--5}. Owing to its kinetic foundation, computational efficiency and intrinsic parallelism, the method has been applied successfully to multiphase flows \refc{6--7}, turbulence \refc{8--9}, combustion \refc{10}, and fluid-structure interaction \refc{11--12}.

Nevertheless, many difficulties in the lattice Boltzmann method arise from the nonlinear collision term, including numerical instability, loss of Galilean invariance in transport coefficients, and the inability to adjust the Prandtl number. Much of the theoretical development over the past several decades has therefore focused on modelling this nonlinear integro-differential term. The simplest collision model is the lattice BGK (LBGK) model developed by Qian et al. \refc{13} from the Boltzmann-BGK equation of Bhatnagar, Gross and Krook \refc{14}. In this model, every component of the non-equilibrium distribution relaxes towards equilibrium with the same relaxation time. Its simplicity and ease of implementation have led to widespread use. The original LBGK model, however, is limited to isothermal flows, has poor numerical stability, and can generate substantial spurious pressure fluctuations. Subsequent collision models have sought to provide an adjustable Prandtl number, improve stability, suppress spurious pressure fluctuations, and recover Galilean-invariant transport coefficients. Although numerical instability in strongly varying flows affects computational fluid dynamics methods in general, it is particularly acute for lattice Boltzmann simulations at high Reynolds or Mach numbers.

D'Humières et al. \refc{15--16} proposed the multiple-relaxation-time (MRT) collision model. In this approach, an orthogonal set of basis vectors is constructed from the discrete velocity set, with the number of basis vectors equal to the number of discrete velocities. The particle distribution function can be used with this basis to calculate moments of different orders, and different relaxation times can be assigned to these moments so that they relax towards their corresponding equilibria. Although the model did not achieve Prandtl-number tunability through multiple relaxation factors, it substantially improved numerical stability. Another collision model that can significantly improve stability and the Galilean invariance of shear viscosity is the cascaded multiple-relaxation-time (cascaded MRT) collision model proposed by Geier et al. \refc{17--18}. In the cascaded MRT model, higher-order moments, namely moments of third order and above, are constructed from lower-order moments and the macroscopic velocity, and each order is relaxed independently.

Lycett-Brown et al. \refc{19--21} started from the one-dimensional case. They first wrote the distribution function explicitly in terms of its zeroth- to second-order moments, and then constructed two- and three-dimensional cascaded MRT collision models by tensor products. Their derivation is more direct and accessible than that of Geier et al. In essence, a cascaded MRT model relaxes the central moments of the distribution function in central-moment space and then transforms the result into raw-moment form through the mathematical relationship between central and raw moments. The relaxation of higher-order moments is thus expressed by lower-order moments and the local macroscopic velocity, which gives rise to the term ``cascaded.'' Other studies of cascaded MRT models include those by Dubois et al. \refc{22--23}, Fei and Luo et al. \refc{24--25}, and De Rosis et al. \refc{26}. However, because of the limitations of the adopted discrete velocities, the Galilean invariance of transport coefficients was not fully resolved in the cascaded MRT model.

In 2015, Geier et al. \refc{27} were inspired by the cumulant method in computational kinetic gas theory proposed by Seeger et al. \refc{28} and introduced the cumulant lattice Boltzmann method (cumulant LBM). Subsequent studies discussed the parametrization of this method \refc{29--30}. In the cumulant multiple-relaxation model, the so-called cumulants are the independently relaxed quantities. The second- and third-order cumulants coincide with the corresponding central moments used in the cascaded collision model; differences first arise at fourth order. Seeger et al. \refc{28} noted in their original study that, in one dimension, their cumulant method in computational kinetic gas theory agrees with the moment method of Grad \refc{31} at least through fifth order, with discrepancies beginning at sixth order. Cumulant collision models have achieved good numerical stability and have been applied to turbulent boundary-layer simulations \refc{32}, aeroacoustic simulations \refc{33}, and free-surface two-phase-flow simulations \refc{34}.

The entropic lattice Boltzmann model proposed by Karlin et al. \refc{35--38} is another collision model with good numerical stability. In this model, the H-theorem is enforced in the discrete molecular-velocity space, and the equilibrium distribution is obtained by iteratively solving an extremum problem. High-order lattices are also constructed as tensor products of one-dimensional lattices. The relaxation factor is no longer fixed but varies dynamically with the flow field, similarly to a large-eddy simulation model. Although the entropic lattice Boltzmann model can markedly improve numerical stability, its computational efficiency offers no advantage because the extremum problem must be solved iteratively at every lattice node and every time step.

Regularized collision models constitute another important route to improved numerical stability. The earliest regularized collision model was proposed by Ladd \refc{39} in 1994 and was successfully applied to sedimenting particle flows \refc{40--41}. Ladd argued that, for solving the weakly compressible Navier--Stokes (N--S) equations, the lattice Boltzmann collision term needs to retain only the relaxation of the second-order stress tensor, including the relaxation of shear and bulk stresses. The relaxation factors of all other higher-order non-hydrodynamic kinetic modes can be set to unity, so that these modes rapidly relax without affecting the N--S equations. The trace part of the second-order stress tensor in the Ladd model is in fact a numerical error caused by insufficient integration accuracy in velocity space; relaxing this term affects numerical stability. In an actual second-order stress tensor, a trace component usually appears only for dense gases or when internal molecular degrees of freedom are excited. The bulk-viscosity relaxation factor should therefore be set to zero, as rigorously discussed by Ladd \refc{41}. In 2005, Latt et al. \refc{42} also proposed a regularized collision model, similar to the Ladd model, and found that regularization substantially improved the stability and accuracy of numerical simulations. They did not, however, discuss separate relaxation of the trace-free and trace parts of the second-order stress tensor, so their model can be regarded as a special case of the Ladd model. Chen et al. \refc{43} discussed, at nearly the same time, whether the Latt-type regularized model could improve the rotational invariance of numerical schemes for high-Knudsen-number flows. Zhang et al. \refc{44}, using the theoretical framework in which Shan et al. \refc{45} expanded the Boltzmann-BGK equation in Hermite polynomials, extended the regularized model to third order, introduced a treatment of the force term, and simulated high-Knudsen-number flows. They argued that the non-equilibrium distribution should be reconstructed in the Hermite-truncated space in a manner analogous to the equilibrium distribution. Their model, however, used raw-moment Hermite space and did not include the first-order term in the treatment of the force.

In 2014, Chen et al. \refc{46} improved the Hermite-expansion MRT model proposed by Shan and Chen in 2007 \refc{47} by replacing the original raw-moment multiple relaxation with multiple relaxation in central-moment space. They found that second-order relaxation remains unchanged, whereas transforming third-order central-moment relaxation into raw-moment form produces an additional correction term related to the local velocity and lower-order moment relaxation. Relaxation of third-order central moments completely removes the previous failure of the thermal conductivity to satisfy Galilean invariance when the Prandtl number differs from one. In 2015, Malaspinas \refc{48} proposed a recursive regularized collision model based on a Hermite expansion and demonstrated a recursive relationship among non-equilibrium Hermite coefficients in the isothermal regime.

This relationship follows from three ingredients. First, the recursive relationship for the equilibrium Hermite coefficients follows by mathematical induction. Second, the Euler equations are obtained from the zeroth-order Chapman--Enskog expansion. Third, the relationship between the non-equilibrium Hermite coefficients and the space-time derivatives of the equilibrium Hermite coefficients follows from the first-order Chapman--Enskog expansion. The author showed that this recursive regularized model has better numerical dissipation and dispersion properties than the original MRT model \refc{15--16}. In 2017, Coreixas et al. \refc{50} extended Malaspinas's model to the compressible thermal regime. Mattila et al. \refc{49} expanded the non-equilibrium distribution in the central-moment Hermite basis and truncated the expansion in central-moment space. For example, discarding fourth- and higher-order terms from the non-equilibrium expansion for compressible thermal flows is equivalent to setting the relaxation factors of fourth- and higher-order non-equilibrium central moments to one. In the isothermal regime, discarding third- and higher-order terms has the analogous interpretation. After transformation to raw-moment space, the non-equilibrium raw-moment Hermite coefficients are exactly the same as those in Malaspinas's recursive regularized model. However, these three regularized models discussed only non-equilibrium Hermite moments and independently relaxed the non-equilibrium raw moments. They did not perform relaxation in central-moment space, so Galilean invariance of the thermal conductivity was not guaranteed.

In 2018, Jacob et al. \refc{51} proposed a hybrid recursive regularized model based on Malaspinas's formulation and modified the reconstruction of the second-order stress tensor. Whereas the original model obtained this tensor from the second moment of the non-equilibrium distribution, the hybrid model also estimated it from central differences of the velocity field. A weighting parameter combined the two estimates and controlled the additional numerical dissipation. Feng et al. \refc{52--54} subsequently incorporated this hybrid model into solvers that treated the mass and momentum equations with the lattice Boltzmann method on D3Q19 and D3Q27 lattices and the energy equation with a finite-volume method. Modifications to the equilibrium distribution and additional source terms minimized the velocity-discretization errors of the standard lattices and enabled transonic and supersonic simulations. In 2019, Shan \refc{55} expanded the non-equilibrium distribution and collision term separately in central-moment space, relaxed each moment order independently, and transformed the collision term back to raw-moment space through the relationship between the two Hermite bases. The resulting recursive collision term retained the advantages of the raw-moment Hermite multiple-relaxation model proposed in 2007 \refc{47}, restored Galilean invariance of the thermal conductivity discussed by Chen et al. \refc{46}, and extended to arbitrary order. Li et al. \refc{56} subsequently performed the expansion and relaxation in temperature-scaled central-moment space. After transformation to raw-moment space, the third- and higher-order collision terms acquire corrections involving velocity, temperature and lower-order relaxed moments relative to direct relaxation in raw-moment space \refc{47}.

Some progress has also been made in boundary-condition treatments for regularized collision models. Niu et al. \refc{57} simulated microscale rarefied-gas flows in 2007 and discussed scattering boundary conditions, relaxation time and regularized collision models. In 2008, Latt et al. \refc{58} developed a regularized boundary condition for flat boundaries based on the reconstruction of the non-equilibrium distribution. In 2011, Malaspinas et al. \refc{59} proposed a general regularized boundary condition for flat boundaries. By establishing an overdetermined system involving the known distribution functions and macroscopic quantities at a boundary node, they solved for the macroscopic moments and reconstructed the non-equilibrium distribution at the boundary. This treatment can be used with standard lattices, such as D2Q9 and D3Q19, and with high-order lattices, such as D2V37 and D3V103. Whether it applies to arbitrary curved surfaces requires further study. In 2017, Wissocq et al. \refc{60} developed regularized characteristic boundary conditions for acoustic simulations. In 2019, Feng et al. \refc{61} discussed arbitrary curved-wall boundary conditions and characteristic open-boundary conditions for compressible flow within a hybrid regularized model.

Regularized collision models have also been applied to multiphase flows. Otomo et al. applied regularized collision models to multicomponent pseudopotential models \refc{62} and phase-field multiphase models \refc{63}. Ba et al. \refc{64} applied a regularized collision model to a color-gradient model. Many representative studies have also used regularized collision models in aeroacoustic simulations and noise prediction, including those by Zhuo et al. \refc{65}, Lew et al. \refc{66}, Avallone et al. \refc{67}, Romani and Casalino \refc{68}, Gendre et al. \refc{69}, and Astoul et al. \refc{70}. Applications of regularized collision models in these fields are not discussed in detail here.

This article reviews the development of regularized collision models. A systematic theoretical framework is used to analyse the high-order regularized collision models developed by the authors and their collaborators, to clarify the theoretical relationships among earlier regularized collision models, and to summarize the essence of regularized multiple-relaxation models. \secref{theory} introduces the theoretical foundations of Hermite expansions and provides the analytical criteria used for regularizing collision operators. \secref{regularization} uses this framework to analyse, organize and compare the classical models that played important roles in the development of regularized collision models. The article concludes with a synthesis of the theoretical findings.

\section{Theoretical foundations}\label{sec:theory}

This section starts from the Boltzmann equation and expands the distribution function in Hermite spectral space. The Maxwell--Boltzmann equilibrium, non-equilibrium distribution, collision term and force term are expanded separately in raw-moment space, central-moment space and temperature-scaled central-moment space, providing the basis for the subsequent review of regularized theoretical models.

We first introduce the Boltzmann equation. It is the governing equation for a rarefied-gas system derived from three assumptions: binary molecular collisions, molecular chaos, and the absence of external-force effects on local collisions. The equation describes the evolution of the particle distribution function in velocity and physical space. Its nonlinear collision term is extremely complicated and difficult to solve directly. As a highly simplified version of the collision term in the Boltzmann equation, the Boltzmann-BGK equation assumes that all kinetic moments relax towards equilibrium at the same rate. Its governing equation is

\begin{equation}
\frac{\partial f}{\partial t}+\xv\!\cdot\!\nabla f+\gv\!\cdot\!\nabla_{\xi}f
\;=\;\Om=-\frac{1}{\tau}\left(f-f^{\mathrm{eq}}\right).
\et{1}
\end{equation}

Here, $f(\boldsymbol{x},\boldsymbol{\xi},t)$ is the phase-space distribution function for particles with molecular velocity $\xv$ at position $\boldsymbol{x}$ and time $t$. The left-hand side of \E{1} describes their transport in space and time, whereas the right-hand side represents local collisions. We choose $c_s=\sqrt{RT_0/m_0}$ as the characteristic speed, where $R$ is the Boltzmann constant, $T_0$ is the reference temperature, and $m_0$ is the molecular mass. The Maxwell--Boltzmann equilibrium distribution in \E{1} is

\begin{equation}
f^{\mathrm{eq}}(\boldsymbol{x},\xv,t)=
\frac{\rho}{(2\pi\theta)^{D/2}}
\exp\!\left[-\frac{(\xv-\uv)^2}{2\theta}\right].
\et{2}
\end{equation}

The temperature $\theta$ is dimensionless and $D$ is the spatial dimension. If $\rho_0$ is taken as the reference density, the dimensionless particle distribution function can be defined as $\overline{f}=f c_s^D/\rho_0$. For simplicity, the overbar on the distribution function is omitted below.

\subsection{Hermite expansion of the particle distribution function}\label{sec:hermite}

Owing to the orthogonality of Hermite polynomials, Grad \refc{31} used them in the kinetic theory of rarefied gases as early as 1949. In this paper, the distribution function is expanded using three Hermite bases with different variables, $\mathcal{H}(\xv)$, $\mathcal{H}(\co)$ and $\mathcal{H}(\boldsymbol{\eta})$, where $\co=\xv-\uv$ and $\boldsymbol{\eta}=\co/\sqrt{\theta}$. The variable of the first basis is the molecular velocity in the absolute coordinate system. The second uses the molecular velocity in a translating coordinate system, and the third uses the local Mach number in that translating system, apart from a factor involving the adiabatic index. The latter variable is dimensionless. The Hermite expansion of a particle distribution function can be viewed as a spectral expansion. In raw-moment space, it is written as

\begin{equation}
f(\boldsymbol{x},\xv,t)=\omega(\xv)\sum_{n=0}^{\infty}
\frac{1}{n!}\,\symbf{a}^{(n)}:\mathcal{H}^{(n)}(\xv).
\et{3}
\end{equation}

The colon on the right-hand side denotes the full contraction of two tensors of order $n$, as it does throughout this paper. The raw-moment Hermite basis is defined by

\begin{equation}
\mathcal{H}^{(n)}(\xv)=\frac{(-1)^n}{\omega(\xv)}\,\nabla_{\xi}^{n}\omega(\xv).
\et{4}
\end{equation}

The corresponding weight function is

\begin{equation}
\omega(\xv)=\frac{1}{(2\pi)^{D/2}}\exp\!\left(-\frac{\xi^2}{2}\right),
\qquad \xi^2=\xv\!\cdot\!\xv.
\et{5}
\end{equation}

Multiplying both sides of \E{3} by $\mathcal{H}^{(n)}(\xv)$ and integrating over molecular-velocity space, the corresponding Hermite coefficient is obtained from the orthogonality of Hermite polynomials:

\begin{equation}
\symbf{a}^{(n)}=\int f \mathcal{H}^{(n)}(\xv)\,\dd\xv.
\et{6}
\end{equation}

\subsection{Truncation of the Hermite expansion and velocity-space discretization}\label{sec:truncation}

In practical numerical calculations, the particle distribution cannot be expanded to infinite order and must be truncated. The truncation order directly affects the range of hydrodynamic macroscopic equations that can be recovered and their accuracy. In practice, the distribution must therefore be projected onto a finite-order Hermite basis. Shan et al. \refc{45} used the Chapman--Enskog expansion to analyse in depth how the truncation order affects the recovered hydrodynamic equations and their accuracy; a brief review follows. Let $\varepsilon$ be a small quantity of the same order as $\mathrm{Kn}$. The distribution function and the temporal and spatial derivatives are expanded perturbatively in powers of $\varepsilon$:

\begin{align}
f&=f^{(0)}+\varepsilon f^{(1)}+\varepsilon^2 f^{(2)}+\cdots,\et{7}\\
\partial_t&=\varepsilon\partial_t^{(0)}+\varepsilon^2\partial_t^{(1)}+\cdots,\et{8}\\
\nabla&=\varepsilon\nabla^{(0)}.\et{9}
\end{align}

With the force omitted and the superscript on $\nabla^{(0)}$ suppressed below, substituting these expressions into the Boltzmann-BGK equation and collecting terms of the same order in $\varepsilon$ gives the following recursive relationship between the $(k+1)$th-order distribution and the space-time derivatives of the $k$th- and lower-order distributions:

\begin{equation}
f^{(k+1)}=-\tau\left(\sum_{m=0}^{k}\partial_t^{(m)}f^{(k-m)}+\xv\!\cdot\!\nabla f^{(k)}\right).
\et{10}
\end{equation}

For example, the first- and second-order recursive relations are

\begin{align}
f^{(1)}&=-\tau\left(\partial_t^{(0)}+\xv\!\cdot\!\nabla\right)f^{(0)},\et{11}\\
f^{(2)}&=-\tau\bigl[(\partial_t^{(0)}+\xv\!\cdot\!\nabla)f^{(1)}\nonumber\\&\qquad+\partial_t^{(1)}f^{(0)}\bigr].\et{12}
\end{align}

Similarly, the $k$th-order distribution can be expanded in a Hermite series:

\begin{equation}
f^{(k)}=\omega(\xv)\sum_{n=0}^{\infty}\frac{1}{n!}
\symbf{a}_k^{(n)}:\mathcal{H}^{(n)}(\xv).
\et{13}
\end{equation}

Using the recursive relation

\begin{equation}
\begin{aligned}
\xi_\alpha\mathcal{H}^{(n)}_{\alpha_1\cdots\alpha_n}(\xv)
&=\mathcal{H}^{(n+1)}_{\alpha\alpha_1\cdots\alpha_n}(\xv)\\
&\quad+\sum_{j=1}^{n}\delta_{\alpha\alpha_j}
\mathcal{H}^{(n-1)}_{\alpha_1\cdots\widehat{\alpha_j}\cdots\alpha_n}(\xv).
\end{aligned}
\et{14}
\end{equation}

where a hat denotes an omitted index, and using the orthogonality of Hermite polynomials gives

\begin{equation}
\int \mathcal{H}^{(n)}(\xv)\left(\xv\!\cdot\!\nabla f\right)\,\dd\xv
=\nabla\!\cdot \symbf{a}^{(n+1)}+n\nabla \symbf{a}^{(n-1)}.
\et{15}
\end{equation}

Substituting \E{13} into \E{10} and using \E{15}, the recursive relation between Hermite coefficients becomes

\begin{equation}
\symbf{a}_{k+1}^{(n)}=-\tau\left(\sum_{m=0}^{k}\partial_t^{(m)}\symbf{a}_{k-m}^{(n)}
+n\nabla \symbf{a}_k^{(n-1)}+\nabla\!\cdot \symbf{a}_k^{(n+1)}\right).
\et{16}
\end{equation}

Here, $k$ denotes the order of the CE expansion and $n$ denotes the order of the Hermite expansion. Each CE order contains a complete Hermite expansion. It follows from \E{16} that the $n$th-order Hermite coefficient in the $(k+1)$th-order CE expansion depends only on Hermite coefficients up to order $n+1$ in the $k$th- and lower-order CE expansions. Therefore, if the Hermite expansion of the equilibrium distribution $f^{(0)}$ is truncated at order $n$, the first $n-k$ Hermite coefficients in the $k$th-order CE expansion are identical to those obtained from the fully expanded equilibrium distribution. The NSF hydrodynamic equations usually solved involve the second-order stress tensor and the third-order heat-flux vector, corresponding to the second- and third-order Hermite coefficients of the non-equilibrium distribution $f^{(1)}$. Exact representation of the second-order stress tensor and third-order heat flux therefore requires the equilibrium Hermite expansion to be truncated at least at third and fourth order, respectively. All Hermite expansions below are consequently truncated at fourth order or below. The recursive relation for Hermite coefficients in \E{16} will be used in the subsequent analysis of regularized models.

Owing to the orthogonality of Hermite polynomials, the low-order moments of the particle distribution can be obtained from a particle distribution represented by a truncated Hermite expansion. The particle distribution can therefore be projected onto the Hilbert space spanned by Hermite polynomials up to order $N$, while preserving exact equality between its first $N$ moments and those of the full distribution. The truncated expansion is

\begin{equation}
f(\boldsymbol{x},\xv,t)\approx f^N(\boldsymbol{x},\xv,t)
=\omega(\xv)\sum_{n=0}^{N}\frac{1}{n!}
\symbf{a}^{(n)}(\boldsymbol{x},t):\mathcal{H}^{(n)}(\xv).
\et{17}
\end{equation}

By orthogonality, the $M$th Hermite coefficients of the infinite and truncated projections are equal provided that $M\leq N$:

\begin{equation}
\symbf{a}^{(M)}(\boldsymbol{x},t)=\int f \mathcal{H}^{(M)}(\xv)\,\dd\xv
=\int f^N \mathcal{H}^{(M)}(\xv)\,\dd\xv.
\et{18}
\end{equation}

After factoring out the Gaussian weight, the integrand on the right-hand side is a polynomial of degree at most $M+N$. It can be evaluated exactly in discrete velocity space by selecting a suitable discrete velocity set \refc{45,71} and applying Gauss-Hermite quadrature:

\begin{equation}
\begin{aligned}
\int\omega(\xv)\frac{f^N(\boldsymbol{x},\xv,t)}{\omega(\xv)}
\mathcal{H}^{(M)}(\xv)\dd\xv\\
=\sum_{i=0}^{b-1}w_i\frac{f^N(\boldsymbol{x},\xv_i,t)}{\omega(\xv_i)}
\mathcal{H}^{(M)}(\xv_i).
\end{aligned}
\et{19}
\end{equation}

The equality in \E{19} is exact when the accuracy $Q$ of the Gauss-Hermite quadrature satisfies $M+N\leq Q$. The discrete-velocity-space particle distribution is defined as $f_i=w_i f^N(\boldsymbol{x},\xv_i,t)/\omega(\xv_i)$. Consequently, the moments of the Hermite-truncated distribution and of its discrete-velocity representation are equivalent to the moments of the continuous distribution. The mathematical requirements are that the moment order does not exceed the Hermite truncation order and that the Gauss-Hermite quadrature accuracy is no less than the sum of the two orders:

\begin{equation}
M\leq N,\qquad M+N\leq Q.
\et{20}
\end{equation}

\subsection{Mathematical relations among different Hermite bases}\label{sec:bases}

For the analysis below, we briefly introduce the transformation relationships among Hermite coefficients in different spaces. Detailed derivations can be found in the appendices of Shan \refc{55,72}, Li et al. \refc{56}, and Shan et al. \refc{72}; only the relationships through fourth order are given here. In actual numerical implementations, the discrete velocities and weights in central-moment space vary with the local macroscopic velocity, whereas those in temperature-scaled central-moment space vary with both the local macroscopic velocity and temperature. This destroys the perfect streaming operation of the lattice Boltzmann method and introduces additional numerical dissipation and dispersion, thereby removing the advantages of the method. Our strategy is therefore to perform collision in central-moment space or temperature-scaled central-moment space, transform the result into raw-moment space through the corresponding mathematical relationship, and finally perform particle streaming in raw-moment space.

The zeroth- to fourth-order raw-moment Hermite bases are

\begin{equation}
\begin{aligned}
\mathcal{H}^{(0)}(\xv)&=1,\\
\mathcal{H}^{(1)}(\xv)&=\xv,\\
\mathcal{H}^{(2)}(\xv)&=\xv\xv-\delv,\\
\mathcal{H}^{(3)}(\xv)&=\xv\xv\xv-3\xv\delv,\\
\mathcal{H}^{(4)}(\xv)&=\xv\xv\xv\xv-6\xv\xv\delv+3\delv\delv.
\end{aligned}
\et{21}
\end{equation}

The mathematical relationship between the central-moment and raw-moment Hermite bases follows from the binomial transformation. Through fourth order,

\begin{equation}
\begin{aligned}
\mathcal{H}^{(0)}(\co)&=1,\\
\mathcal{H}^{(1)}(\co)&=\mathcal{H}^{(1)}(\xv)-\uv,\\
\mathcal{H}^{(2)}(\co)&=\mathcal{H}^{(2)}(\xv)-2\mathcal{H}^{(1)}(\xv)\uv+\uv\uv,\\
\mathcal{H}^{(3)}(\co)&=\mathcal{H}^{(3)}(\xv)-3\mathcal{H}^{(2)}(\xv)\uv+3\mathcal{H}^{(1)}(\xv)\uv\uv-\uv\uv\uv,\\
\mathcal{H}^{(4)}(\co)&=\mathcal{H}^{(4)}(\xv)-4\mathcal{H}^{(3)}(\xv)\uv
 +6\mathcal{H}^{(2)}(\xv)\uv\uv\\
&\quad-4\mathcal{H}^{(1)}(\xv)\uv\uv\uv+\uv\uv\uv\uv.
\end{aligned}
\et{22}
\end{equation}

The mathematical relationship between the temperature-scaled central-moment and raw-moment Hermite bases can be proved by mathematical induction. Through fourth order,

\begin{equation}
\begin{aligned}
\mathcal{H}^{(0)}(\boldsymbol{\eta})&=1,\\
\mathcal{H}^{(1)}(\boldsymbol{\eta})&=\theta^{-1/2}\left[\mathcal{H}^{(1)}(\xv)-\uv\right],\\
\mathcal{H}^{(2)}(\boldsymbol{\eta})&=\theta^{-1}\left[\mathcal{H}^{(2)}(\xv)-2\mathcal{H}^{(1)}(\xv)\uv+\uv_1\right],\\
\mathcal{H}^{(3)}(\boldsymbol{\eta})&=\theta^{-3/2}[\mathcal{H}^{(3)}(\xv)-3\mathcal{H}^{(2)}(\xv)\uv\\
&\quad+3\mathcal{H}^{(1)}(\xv)\uv_1-\uv\uv_2],\\
\mathcal{H}^{(4)}(\boldsymbol{\eta})&=\theta^{-2}[\mathcal{H}^{(4)}(\xv)-4\mathcal{H}^{(3)}(\xv)\uv
 +6\mathcal{H}^{(2)}(\xv)\uv_1\\
&\quad-4\mathcal{H}^{(1)}(\xv)\uv\uv_2+\uv\uv\uv_2\\
&\quad-3(\theta-1)\uv_1\delv].
\end{aligned}
\et{23}
\end{equation}

Here, $\uv_1\equiv\uv\uv-(\theta-1)\delv$ and $\uv_2\equiv\uv\uv-3(\theta-1)\delv$. In \E{21}, \E{22}, \E{23}, and the higher-order symmetric tensors used below, a coefficient preceding a tensor denotes the number of permutations. For example,
$4\mathcal{H}^{(1)}(\xv)u u u=\mathcal{H}_\alpha u_\beta u_\gamma u_\delta+\mathcal{H}_\beta u_\gamma u_\delta u_\alpha+\mathcal{H}_\gamma u_\delta u_\alpha u_\beta+\mathcal{H}_\delta u_\alpha u_\beta u_\gamma$, where the indices $\alpha$, $\beta$, $\gamma$ and $\delta$ can take the three-dimensional coordinate components $x$, $y$ and $z$.

\subsection{Hermite expansions and coefficient evaluation in different spaces}\label{sec:coefficients}

Existing regularized models differ mainly in three respects. First, they use different Hermite bases, namely expansions in raw-moment, central-moment or temperature-scaled central-moment space. Second, they use different truncation orders for the Hermite expansion of the distribution function, especially for the non-equilibrium distribution. Third, they differ in the space in which the collision term is expanded and in the tensor-moment components to which relaxation is applied. To review these regularized collision models systematically, we expand the equilibrium distribution, non-equilibrium distribution, collision term and force term in all three Hermite bases and analyse them within a common framework.

\subsubsection{Hermite expansion of the equilibrium distribution}\label{sec:equilibrium}

Using the definition of the weight function in \E{5}, the Maxwell--Boltzmann equilibrium distribution can be written as

\begin{equation}
f^{(0)}=\frac{\rho}{(\sqrt{\theta})^D}
\omega\!\left(\frac{\xv-\uv}{\sqrt{\theta}}\right).
\et{24}
\end{equation}

Similarly, the equilibrium distribution can be approximated by its projection onto the Hilbert space spanned by Hermite polynomials up to order $N$:

\begin{equation}
f^{(0)}(\boldsymbol{x},\xv,t)\approx f^{(0),N}(\boldsymbol{x},\xv,t)
=\omega(\xv)\sum_{n=0}^{N}\frac{1}{n!}\symbf{a}_0^{(n)}:\mathcal{H}^{(n)}(\xv).
\et{25}
\end{equation}

The raw-moment Hermite coefficients of the equilibrium distribution are calculated from

\begin{equation}
\symbf{a}_0^{(n)}=\int f^{(0)}\mathcal{H}^{(n)}(\xv)\,\dd\xv.
\et{26}
\end{equation}

Substitution of the Maxwell--Boltzmann equilibrium distribution into \E{26}, together with Hermite orthogonality, gives the zeroth- to fourth-order raw-moment coefficients \refc{45}:

\begin{equation}
\begin{aligned}
\symbf{a}_0^{(0)}&=\rho,\\
\symbf{a}_0^{(1)}&=\rho\uv,\\
\symbf{a}_0^{(2)}&=\rho\left[\uv\uv+(\theta-1)\delv\right],\\
\symbf{a}_0^{(3)}&=\rho\left[\uv\uv\uv+3(\theta-1)\delv\uv\right],\\
\symbf{a}_0^{(4)}&=\rho\left[\uv\uv\uv\uv+6(\theta-1)\delv\uv\uv
 +3(\theta-1)^2\delv\delv\right].
\end{aligned}
\et{27}
\end{equation}

If the equilibrium distribution is expanded in central-moment space using $\mathcal{H}^{(n)}(\xv-\uv)$, then

\begin{equation}
\begin{aligned}
f^{(0)}(\boldsymbol{x},\xv,t)&\approx f^{(0),N}(\boldsymbol{x},\xv,t)\\
&=\omega(\co)\sum_{n=0}^{N}\frac{1}{n!}\symbf{b}_0^{(n)}:\mathcal{H}^{(n)}(\co),\\
&\qquad \co=\xv-\uv.
\end{aligned}
\et{28}
\end{equation}

The corresponding central-moment Hermite coefficients follow from orthogonality:

\begin{equation}
\symbf{b}_0^{(n)}=\int f^{(0)}\mathcal{H}^{(n)}(\co)\,\dd\co.
\et{29}
\end{equation}

Using the transformation in \E{22}, the equilibrium central-moment coefficients through fourth order are

\begin{equation}
\begin{aligned}
\symbf{b}_0^{(0)}&=\rho, & \symbf{b}_0^{(1)}&=0, & \symbf{b}_0^{(2)}&=\rho(\theta-1)\delv,\\
\symbf{b}_0^{(3)}&=0, & \symbf{b}_0^{(4)}&=3\rho(\theta-1)^2\delv\delv.
\end{aligned}
\et{30}
\end{equation}

The odd-order central-moment Hermite coefficients of the equilibrium distribution vanish, and the coefficients no longer contain the local macroscopic velocity. Thus, these moments are independent of uniform translations of the reference frame.

If the equilibrium distribution is expanded in temperature-scaled central-moment space using $\mathcal{H}^{(n)}(\boldsymbol{\eta})$, then

\begin{equation}
\begin{aligned}
f^{(0)}(\boldsymbol{x},\xv,t)&\approx f^{(0),N}(\boldsymbol{x},\xv,t)\\
&=\omega(\boldsymbol{\eta})\sum_{n=0}^{N}\frac{1}{n!}\symbf{d}_0^{(n)}:\mathcal{H}^{(n)}(\boldsymbol{\eta}),\\
&\qquad \boldsymbol{\eta}=\frac{\xv-\uv}{\sqrt{\theta}}.
\end{aligned}
\et{31}
\end{equation}

The temperature-scaled central-moment Hermite coefficients are

\begin{equation}
\symbf{d}_0^{(n)}=\int f^{(0)}\mathcal{H}^{(n)}(\boldsymbol{\eta})\frac{1}{(\sqrt{\theta})^D}\,\dd\xv.
\et{32}
\end{equation}

Because $\dd\boldsymbol{\eta}=\dd\xv/(\sqrt{\theta})^D$, the equilibrium expression in \E{24} and Hermite orthogonality show that only the zeroth-order integral is nonzero:

\begin{equation}
\symbf{d}_0^{(0)}=\frac{\rho}{(\sqrt{\theta})^D}.
\et{33}
\end{equation}

\subsubsection{Hermite expansion of the non-equilibrium distribution}\label{sec:non-equilibrium}

From this subsection onward, $f^{(1)}$ denotes the non-equilibrium part $f-f^{(0)}$; its leading Chapman--Enskog approximation is used in \E{37}. The non-equilibrium distribution can likewise be approximated by its projection onto the Hilbert space spanned by Hermite polynomials up to order $N$. With the raw-moment Hermite basis,

\begin{equation}
f^{(1)}(\boldsymbol{x},\xv,t)\approx f^{(1),N}(\boldsymbol{x},\xv,t)
=\omega(\xv)\sum_{n=0}^{N}\frac{1}{n!}\symbf{a}_1^{(n)}:\mathcal{H}^{(n)}(\xv).
\et{34}
\end{equation}

The raw-moment Hermite coefficients are

\begin{equation}
\symbf{a}_1^{(n)}=\int f^{(1)}\mathcal{H}^{(n)}(\xv)\,\dd\xv.
\et{35}
\end{equation}

The zeroth- to third-order raw-moment Hermite coefficients of the non-equilibrium distribution are

\begin{equation}
\begin{aligned}
\symbf{a}_1^{(0)}&=\int f^{(1)}\,\dd\xv=0,\\
\symbf{a}_1^{(1)}&=\int f^{(1)}\xv\,\dd\xv=0,\\
\symbf{a}_1^{(2)}&=\int f^{(1)}(\xv\xv-\delv)\,\dd\xv\approx\boldsymbol{\Pi}^{(1)},\\
\symbf{a}_1^{(3)}&=\int f^{(1)}(\xv\xv\xv-3\delv\xv)\,\dd\xv\approx\boldsymbol{Q}^{(1)}.
\end{aligned}
\et{36}
\end{equation}

The second- and third-order non-equilibrium coefficients can be obtained by the Chapman--Enskog expansion \refc{73}:

\begin{equation}
\begin{aligned}
\Pi^{(1)}_{\alpha\beta}&=-\tau\rho\theta\left(\partial_\alpha u_\beta+\partial_\beta u_\alpha
-\frac{2}{D}\delta_{\alpha\beta}\partial_\gamma u_\gamma\right),\\
Q^{(1)}_{\alpha\beta\gamma}&=u_\alpha a^{(2)}_{1,\beta\gamma}+u_\beta a^{(2)}_{1,\alpha\gamma}+u_\gamma a^{(2)}_{1,\alpha\beta}\\
&\quad-\tau\rho\theta\left(\delta_{\alpha\beta}\partial_\gamma\theta+\delta_{\beta\gamma}\partial_\alpha\theta+\delta_{\alpha\gamma}\partial_\beta\theta\right).
\end{aligned}
\et{37}
\end{equation}

If the non-equilibrium distribution is expanded in the central-moment Hermite basis, then

\begin{equation}
f^{(1)}(\boldsymbol{x},\xv,t)\approx f^{(1),N}(\boldsymbol{x},\xv,t)
=\omega(\co)\sum_{n=0}^{N}\frac{1}{n!}\symbf{b}_1^{(n)}:\mathcal{H}^{(n)}(\co).
\et{38}
\end{equation}

The central-moment Hermite coefficients are

\begin{equation}
\symbf{b}_1^{(n)}=\int f^{(1)}\mathcal{H}^{(n)}(\co)\,\dd\co.
\et{39}
\end{equation}

Using the relationship between the central-moment and raw-moment Hermite bases gives the explicit expressions through fourth order:

\begin{align}
\symbf{b}_1^{(0)}&=0,\et{40}\\
\symbf{b}_1^{(1)}&=0,\et{41}\\
\symbf{b}_1^{(2)}&=\symbf{a}_1^{(2)},\et{42}\\
\symbf{b}_1^{(3)}&=\symbf{a}_1^{(3)}-3\uv \symbf{a}_1^{(2)},\et{43}\\
\symbf{b}_1^{(4)}&=\symbf{a}_1^{(4)}-4\uv \symbf{a}_1^{(3)}+6\uv\uv \symbf{a}_1^{(2)}.\et{44}
\end{align}

Conversely, the raw-moment non-equilibrium Hermite coefficients can be expressed in terms of their central-moment counterparts:

\begin{equation}
\begin{aligned}
\symbf{a}_1^{(0)}&=0,\\
\symbf{a}_1^{(1)}&=0,\\
\symbf{a}_1^{(2)}&=\symbf{b}_1^{(2)},\\
\symbf{a}_1^{(3)}&=\symbf{b}_1^{(3)}+3\uv \symbf{b}_1^{(2)},\\
\symbf{a}_1^{(4)}&=\symbf{b}_1^{(4)}+4\uv \symbf{b}_1^{(3)}+6\uv\uv \symbf{b}_1^{(2)}.
\end{aligned}
\et{45}
\end{equation}

More generally, if the non-equilibrium distribution is expanded in temperature-scaled central-moment space,

\begin{equation}
f^{(1)}(\boldsymbol{x},\xv,t)\approx f^{(1),N}(\boldsymbol{x},\xv,t)
=\omega(\boldsymbol{\eta})\sum_{n=0}^{N}\frac{1}{n!}\symbf{d}_1^{(n)}:\mathcal{H}^{(n)}(\boldsymbol{\eta}).
\et{46}
\end{equation}

The corresponding coefficients are

\begin{equation}
\symbf{d}_1^{(n)}=\int f^{(1)}\mathcal{H}^{(n)}(\boldsymbol{\eta})\frac{1}{(\sqrt{\theta})^D}\,\dd\xv.
\et{47}
\end{equation}

The transformation relationship gives, through fourth order,

\begin{equation}
\begin{aligned}
\symbf{d}_1^{(0)}&=0, &\symbf{d}_1^{(1)}&=0,\\
\symbf{d}_1^{(2)}&=\left(\frac{1}{\sqrt{\theta}}\right)^{D+2}\symbf{a}_1^{(2)},\\
\symbf{d}_1^{(3)}&=\left(\frac{1}{\sqrt{\theta}}\right)^{D+3}
\left(\symbf{a}_1^{(3)}-3\uv \symbf{a}_1^{(2)}\right),\\
\symbf{d}_1^{(4)}&=\left(\frac{1}{\sqrt{\theta}}\right)^{D+4}
\left[\symbf{a}_1^{(4)}-4\uv \symbf{a}_1^{(3)}\right.\\
&\left.\quad+6\left(\uv\uv-(\theta-1)\delv\right)\symbf{a}_1^{(2)}\right].
\end{aligned}
\et{48}
\end{equation}

The inverse relationship is

\begin{equation}
\begin{aligned}
\symbf{a}_1^{(0)}&=0, &\symbf{a}_1^{(1)}&=0,\\
\symbf{a}_1^{(2)}&=(\sqrt{\theta})^{D+2}\symbf{d}_1^{(2)},\\
\symbf{a}_1^{(3)}&=(\sqrt{\theta})^{D+3}\symbf{d}_1^{(3)}+3\uv(\sqrt{\theta})^{D+2}\symbf{d}_1^{(2)},\\
\symbf{a}_1^{(4)}&=(\sqrt{\theta})^{D+4}\symbf{d}_1^{(4)}+4\uv(\sqrt{\theta})^{D+3}\symbf{d}_1^{(3)}\\
&\quad+6\left[\uv\uv+(\theta-1)\delv\right](\sqrt{\theta})^{D+2}\symbf{d}_1^{(2)}.
\end{aligned}
\et{49}
\end{equation}

\subsubsection{Hermite expansion of the force term}\label{sec:force}

The force term in \E{1} is in continuous form. We now discuss its representation in the Hilbert space spanned by Hermite polynomials up to order $N$. The force term is $F(\xv)=-\gv\!\cdot\!\nabla_{\xi}f^N$. Using the Rodrigues representation of the Hermite basis in \E{4}, the derivative of the distribution with respect to particle velocity is

\begin{equation}
\begin{aligned}
\partial_{\xi_\alpha} f^N
&=\sum_{n=0}^{N}\frac{a^{(n)}_{\beta_1\cdots\beta_n}}{n!}
\partial_{\xi_\alpha}\bigl[\omega(\xv)
\mathcal{H}^{(n)}_{\beta_1\cdots\beta_n}(\xv)\bigr]\\
&=-\omega(\xv)\sum_{n=0}^{N}\frac{a^{(n)}_{\beta_1\cdots\beta_n}}{n!}
\mathcal{H}^{(n+1)}_{\alpha\beta_1\cdots\beta_n}(\xv)\\
&=-\omega(\xv)\sum_{n=1}^{N+1}\frac{a^{(n-1)}_{\beta_1\cdots\beta_{n-1}}}{(n-1)!}
\mathcal{H}^{(n)}_{\alpha\beta_1\cdots\beta_{n-1}}(\xv).
\end{aligned}
\et{50}
\end{equation}

Thus, the truncated Hermite expansion of the force term is

\begin{equation}
F(\xv)\approx\omega(\xv)\sum_{n=1}^{N+1}\frac{1}{n!}
\left(n\gv \symbf{a}^{(n-1)}\right):\mathcal{H}^{(n)}(\xv).
\et{51}
\end{equation}

Repeated component indices in \E{50} are summed. Here, $n\gv \symbf{a}^{(n-1)}$ denotes the symmetrized dyadic product of $\symbf{a}^{(n-1)}$ and $\gv$ and is a symmetric tensor. For example, the third-order form is $3\gv \symbf{a}^{(2)}=g_\alpha a_{\beta\gamma}+g_\beta a_{\gamma\alpha}+g_\gamma a_{\alpha\beta}$. The force-term expansions in raw-moment, central-moment and temperature-scaled central-moment space are respectively

\begin{equation}
\begin{aligned}
F(\xv)&\approx\omega(\xv)\sum_{n=1}^{N+1}\frac{1}{n!}\,n\gv\left(\symbf{a}_0^{(n-1)}+\symbf{a}_1^{(n-1)}\right):\mathcal{H}^{(n)}(\xv),\\
F(\xv)&\approx\omega(\co)\sum_{n=1}^{N+1}\frac{1}{n!}\,n\gv\left(\symbf{b}_0^{(n-1)}+\symbf{b}_1^{(n-1)}\right):\mathcal{H}^{(n)}(\co),\\
F(\xv)&\approx\frac{\omega(\boldsymbol{\eta})}{\sqrt{\theta}}
\sum_{n=1}^{N+1}\frac{1}{n!}\,n\gv\left(\symbf{d}_0^{(n-1)}+\symbf{d}_1^{(n-1)}\right):\mathcal{H}^{(n)}(\boldsymbol{\eta}).
\end{aligned}
\et{52}
\end{equation}

Here, the relationships $\nabla_\xi=\nabla_c$ and $\nabla_\xi=\theta^{-1/2}\nabla_\eta$ have been used. Let $N=3$. As discussed below, the non-equilibrium expansion is one order lower than the equilibrium expansion and the contribution of higher-order terms is small. The explicit truncated force-term expansions are given below; the labels beneath the terms indicate their Hermite order; the equilibrium coefficients are given in \E{27}:

\begin{equation}
\begin{aligned}
\frac{F(\xv)}{\omega(\xv)}\approx{}&
\underbrace{\rho\gv\!\cdot\!\xv}_{\text{1st}}
+\underbrace{\rho[(\gv\!\cdot\!\xv)(\xv\!\cdot\!\uv)-\gv\!\cdot\!\uv]}_{\text{2nd}}\\
&+\underbrace{\frac{3\gv(\symbf{a}_0^{(2)}+\symbf{a}_1^{(2)}):\mathcal{H}^{(3)}(\xv)}{6}}_{\text{3rd}}\\
&+\underbrace{\frac{4\gv(\symbf{a}_0^{(3)}+\symbf{a}_1^{(3)}):\mathcal{H}^{(4)}(\xv)}{24}}_{\text{4th}}.
\end{aligned}
\et{53}
\end{equation}

\begin{equation}
\begin{aligned}
\frac{F(\xv)}{\omega(\co)}\approx{}&\underbrace{\rho\gv\!\cdot\!\co}_{\text{1st}}
+\underbrace{0}_{\text{2nd}}\\
&+\underbrace{\frac{3\gv[\rho(\theta-1)\delv+\symbf{a}_1^{(2)}]:\mathcal{H}^{(3)}(\co)}{6}}_{\text{3rd}}\\
&+\underbrace{\frac{4\gv(\symbf{a}_1^{(3)}-3\uv\symbf{a}_1^{(2)}):\mathcal{H}^{(4)}(\co)}{24}}_{\text{4th}}.
\end{aligned}
\et{54}
\end{equation}

\begin{equation}
\begin{aligned}
\frac{F(\xv)}{\omega(\boldsymbol{\eta})}\approx{}&
\underbrace{\frac{\rho\gv\!\cdot\!\boldsymbol{\eta}}{\theta^{(D+1)/2}}}_{\text{1st}}
+\underbrace{0}_{\text{2nd}}\\
&+\underbrace{\frac{3\gv\symbf{a}_1^{(2)}:\mathcal{H}^{(3)}(\boldsymbol{\eta})}{6\theta^{(D+3)/2}}}_{\text{3rd}}\\
&+\underbrace{\frac{4\gv(\symbf{a}_1^{(3)}-3\uv\symbf{a}_1^{(2)}):\mathcal{H}^{(4)}(\boldsymbol{\eta})}{24\theta^{(D+4)/2}}}_{\text{4th}}.
\end{aligned}
\et{55}
\end{equation}

Equation~\eqref{eq:53} is the form derived by Martys et al. \refc{74}. The contribution of non-equilibrium Hermite moments to the force term is very small and can often be neglected in continuum-flow calculations. The widely used force treatment of He et al. \refc{75} replaces the distribution in \E{50} with its equilibrium counterpart. Analogously to the distribution function, the force is discretized in velocity space as $F_i=w_iF(\xv_i)/\omega(\xv_i)$.

In \Es{56}{64}, the non-equilibrium contribution to the force is neglected. Let $\Fv=\rho\gv$. The raw-moment Hermite coefficients of the force-term expansion are

\begin{equation}
\begin{aligned}
\symbf{a}_F^{(1)}&=\Fv,\\
\symbf{a}_F^{(2)}&=2\Fv\uv,\\
\symbf{a}_F^{(3)}&=3\Fv\left[\uv\uv+(\theta-1)\delv\right],\\
\symbf{a}_F^{(4)}&=4\Fv\left[\uv\uv\uv+3(\theta-1)\uv\delv\right].
\end{aligned}
\et{56}
\end{equation}

The central-moment Hermite coefficients are

\begin{equation}
\symbf{b}_F^{(1)}=\Fv.
\et{57}
\end{equation}

\begin{equation}
\symbf{b}_F^{(2)}=0.
\et{58}
\end{equation}

\begin{equation}
\symbf{b}_F^{(3)}=3\Fv\delv(\theta-1).
\et{59}
\end{equation}

\begin{equation}
\symbf{b}_F^{(4)}=0.
\et{60}
\end{equation}

The temperature-scaled central-moment Hermite coefficients are

\begin{equation}
\symbf{d}_F^{(1)}=\frac{\Fv}{(\sqrt{\theta})^{D+1}},\qquad
\symbf{d}_F^{(2)}=0,\qquad \symbf{d}_F^{(3)}=0,\qquad \symbf{d}_F^{(4)}=0.
\et{61}
\end{equation}

\Es{56}{61} show that, when the contribution of the non-equilibrium distribution is ignored, the force-term Hermite coefficients in each space are consistent with the corresponding equilibrium Hermite coefficients. In particular, only the first-order coefficient is nonzero in temperature-scaled central-moment space, so only the first Hermite mode is forced directly in this space. In raw-moment space, the same force also supplies the mechanical-work contribution to the total-energy equation.

Retaining only terms through second order in Eqs.~\eqref{eq:56}--\eqref{eq:58} yields

\begin{equation}
F(\xv)\approx\omega(\xv)\rho\left[\gv\!\cdot\!\xv+(\gv\!\cdot\!\xv)(\xv\!\cdot\!\uv)-\gv\!\cdot\!\uv\right].
\et{62}
\end{equation}

\begin{equation}
F(\xv)\approx\omega(\xv-\uv)\rho\gv\!\cdot\!(\xv-\uv)
=\gv\!\cdot\!(\xv-\uv)f^{\mathrm{eq}}
=-\gv\!\cdot\!\nabla_{\xi- u}f^{\mathrm{eq}}.
\et{63}
\end{equation}

\begin{equation}
\begin{aligned}
F(\xv)&\approx\frac{\rho\omega(\boldsymbol{\eta})}{\theta^{(D+1)/2}}
\gv\!\cdot\!\boldsymbol{\eta}\\
&=\frac{\gv\!\cdot\!\boldsymbol{\eta}}{\sqrt{\theta}}f^{\mathrm{eq}}
=\frac{\gv\!\cdot\!\co}{\theta}f^{\mathrm{eq}}
=-\gv\!\cdot\!\nabla_\xi f^{\mathrm{eq}}.
\end{aligned}
\et{64}
\end{equation}

The derivation uses the Maxwell--Boltzmann equilibrium distribution, with the isothermal assumption applied in deriving \E{63}. Discretizing \E{62} in velocity space and accounting for lattice effects recovers the force scheme derived by Guo et al. \refc{76} in 2002; its discrete form is discussed below. Equation~\eqref{eq:63} is the force scheme proposed by He et al. \refc{75} in 1998, and \E{64} reduces to \E{63} when $\theta=1$.

\subsubsection{Hermite expansion of the collision term}\label{sec:collision}

We introduce two projection approaches for the collision term. The first expands the collision term using Hermite bases in raw-moment, central-moment and temperature-scaled central-moment spaces, namely $\mathcal{H}^{(n)}(\xv)$, $\mathcal{H}^{(n)}(\co)$ and $\mathcal{H}^{(n)}(\boldsymbol{\eta})$. The second uses monomial power bases in the same three spaces, namely $\xv^n$, $\co^n$ and $\boldsymbol{\eta}^n$. The second approach was adopted by Chen et al. \refc{46} and Shan \refc{55}. To facilitate comparison and clarify the relationships among the models, both approaches are discussed.

Using a Hermite basis, the right-hand side of the Boltzmann-BGK equation can be expanded as

\begin{equation}
\Om=\omega(\boldsymbol{\varsigma})\sum_{n=2}^{N}\frac{1}{n!}
\symbf{p}_{\Omega}^{(n)}:\mathcal{H}^{(n)}(\boldsymbol{\varsigma}),
\et{65}
\end{equation}

where $\symbf{p}_{\Omega}^{(n)}=\symbf{a}_{\Omega}^{(n)}$, $\symbf{b}_{\Omega}^{(n)}$ or $\symbf{d}_{\Omega}^{(n)}$ corresponds to the collision-term Hermite coefficient in one of the three spaces. For convenience, the non-equilibrium distribution expanded in the three spaces is written again as

\begin{equation}
\begin{aligned}
f^{(1)}&=\omega(\boldsymbol{\varsigma})\sum_{n=2}^{N}\frac{1}{n!}
\symbf{q}_1^{(n)}:\mathcal{H}^{(n)}(\boldsymbol{\varsigma}),\\
&\qquad \symbf{q}_1^{(n)}=\symbf{a}_1^{(n)},\;\symbf{b}_1^{(n)}\;\text{or}\;\symbf{d}_1^{(n)}.
\end{aligned}
\et{66}
\end{equation}

where the variable $\boldsymbol{\varsigma}$ can be any of $\xv$, $\co$ and $\boldsymbol{\eta}$. The transformation relationships between the collision-term Hermite coefficients and the non-equilibrium Hermite coefficients are the same in the different spaces and will be used directly. As pointed out by Chen et al. \refc{46}, a multiple-relaxation collision term must satisfy certain constraints to preserve Galilean invariance of the transport coefficients. The relaxation of the moments of the non-equilibrium distribution must be equivalent, at the order under consideration, to the relaxation of the original collision term:

\begin{equation}
\int \Omega\,\boldsymbol{\varsigma}^{n}\,\dd\boldsymbol{\varsigma}
=-\frac{1}{\tau_n}\int f^{(1)}\boldsymbol{\varsigma}^{n}\,\dd\boldsymbol{\varsigma}.
\et{67}
\end{equation}

Here, $\omega_n=1/\tau_n$ denotes the continuous relaxation rate. Multiple relaxation times are used rather than the single relaxation factor in the original Boltzmann-BGK equation. Different relaxation factors are applied to moments of different orders, giving an extended model. In our work, the constraint is extended to relaxation of the Hermite moments of the non-equilibrium distribution. This relaxation must likewise be equivalent, at the order under consideration, to relaxation of the original collision term:

\begin{equation}
\int \Omega \mathcal{H}^{(n)}(\boldsymbol{\varsigma})\,\dd\boldsymbol{\varsigma}
=-\frac{1}{\tau_n}\int f^{(1)}\mathcal{H}^{(n)}(\boldsymbol{\varsigma})\,\dd\boldsymbol{\varsigma}.
\et{68}
\end{equation}

We first discuss the constraint in \E{67}. The relationship between $\boldsymbol{\varsigma}^{n}$ and the first few Hermite bases follows from \E{21}; the general relationship is given in Shan \refc{55}. Through fourth order,

\begin{equation}
\begin{aligned}
1&=\mathcal{H}^{(0)}(\boldsymbol{\varsigma}),\\
\boldsymbol{\varsigma}&=\mathcal{H}^{(1)}(\boldsymbol{\varsigma}),\\
\boldsymbol{\varsigma}\boldsymbol{\varsigma}&=\mathcal{H}^{(2)}(\boldsymbol{\varsigma})+\delv \mathcal{H}^{(0)}(\boldsymbol{\varsigma}),\\
\boldsymbol{\varsigma}^3&=\mathcal{H}^{(3)}(\boldsymbol{\varsigma})+3\delv \mathcal{H}^{(1)}(\boldsymbol{\varsigma}),\\
\boldsymbol{\varsigma}^4&=\mathcal{H}^{(4)}(\boldsymbol{\varsigma})+6\mathcal{H}^{(2)}(\boldsymbol{\varsigma})\delv
+3\delv\delv \mathcal{H}^{(0)}(\boldsymbol{\varsigma}).
\end{aligned}
\et{69}
\end{equation}

Substituting \E{69} and the Hermite expansions of the collision term and non-equilibrium distribution into \E{67}, and using Hermite orthogonality, gives

\begin{equation}
\begin{aligned}
\symbf{p}_{\Omega}^{(2)}&=-\omega_2\symbf{q}_1^{(2)},\\
\symbf{p}_{\Omega}^{(3)}&=-\omega_3\symbf{q}_1^{(3)},\\
\symbf{p}_{\Omega}^{(4)}+6\delv \symbf{p}_{\Omega}^{(2)}
&=-\omega_4\left(\symbf{q}_1^{(4)}+6\delv \symbf{q}_1^{(2)}\right).
\end{aligned}
\et{70}
\end{equation}

Let $\symbf{p}_{\Omega}^{(n)}=\symbf{b}_{\Omega}^{(n)}$ and $\symbf{q}_1^{(n)}=\symbf{b}_1^{(n)}$, corresponding to expansion in central-moment space. Substitution of the central-moment and raw-moment coefficient relationships in Eqs. \eqref{eq:40}--\eqref{eq:45} gives

\begin{equation}
\begin{aligned}
\symbf{a}_{\Omega}^{(2)}&=-\omega_2 \symbf{b}_1^{(2)}=-\omega_2\symbf{a}_1^{(2)},\\
\symbf{a}_{\Omega}^{(3)}-3\uv \symbf{a}_{\Omega}^{(2)}
&=-\omega_3\symbf{b}_1^{(3)}=-\omega_3\left(\symbf{a}_1^{(3)}-3\uv \symbf{a}_1^{(2)}\right),\\
\symbf{a}_{\Omega}^{(4)}-4\uv \symbf{a}_{\Omega}^{(3)}\\
&\quad+6(\uv\uv+\delv)\symbf{a}_{\Omega}^{(2)}
=-\omega_4\left(\symbf{b}_1^{(4)}+6\delv \symbf{b}_1^{(2)}\right)\\
&=-\omega_4[\symbf{a}_1^{(4)}-4\uv \symbf{a}_1^{(3)}\\
&\quad+6(\uv\uv+\delv)\symbf{a}_1^{(2)}].
\end{aligned}
\et{71}
\end{equation}

Rearranging \E{71}, the collision-term coefficients are

\begin{align}
\symbf{a}_{\Omega}^{(2)}&=-\omega_2\symbf{a}_1^{(2)},\et{72}\\
\symbf{a}_{\Omega}^{(3)}&=-\omega_3\symbf{a}_1^{(3)}+3(\omega_3-\omega_2)\uv \symbf{a}_1^{(2)},\et{73}\\
\symbf{a}_{\Omega}^{(4)}&=-\omega_4\symbf{a}_1^{(4)}+4(\omega_4-\omega_3)\uv \symbf{a}_1^{(3)}\nonumber\\
&\quad-6(\omega_4+\omega_2-2\omega_3)\uv\uv \symbf{a}_1^{(2)}\nonumber\\
&\quad-6(\omega_4-\omega_2)\delv \symbf{a}_1^{(2)}.\et{74}
\end{align}

If $\symbf{p}_{\Omega}^{(n)}=\symbf{d}_{\Omega}^{(n)}$ and $\symbf{q}_1^{(n)}=\symbf{d}_1^{(n)}$, the same algebra gives exactly the same collision-term Hermite coefficients. Thus, under the constraint in \E{67}, using powers of $\co$ or of $\boldsymbol{\eta}$ gives the same collision-term Hermite coefficients; this statement does not extend to powers of the unshifted velocity $\xv$. The second- and third-order terms in \E{72} and \E{73} are equivalent to the collision form derived by Chen et al. \refc{46}. The second term on the right-hand side of \E{73} is a third-order relaxation correction to the multiple-relaxation model proposed by Shan et al. \refc{47} in 2007. In 2019, Shan \refc{55} extended the work of Chen et al. to arbitrary orders and truncated it at fourth order, giving \E{74}.

We next derive the collision-term Hermite coefficients from the constraint in \E{68}. Substituting the Hermite expansions of the collision term and non-equilibrium distribution into \E{68}, and using Hermite orthogonality, yields

\begin{equation}
\begin{aligned}
\symbf{p}_{\Omega}^{(2)}&=-\omega_2\symbf{q}_1^{(2)},\\
\symbf{p}_{\Omega}^{(3)}&=-\omega_3\symbf{q}_1^{(3)},\\
\symbf{p}_{\Omega}^{(4)}&=-\omega_4\symbf{q}_1^{(4)}.
\end{aligned}
\et{75}
\end{equation}

For $\symbf{p}_{\Omega}^{(n)}=\symbf{a}_{\Omega}^{(n)}$ and $\symbf{q}_1^{(n)}=\symbf{a}_1^{(n)}$, the raw-moment-space collision form is

\begin{equation}
\begin{aligned}
\symbf{a}_{\Omega}^{(2)}&=-\omega_2\symbf{a}_1^{(2)},\\
\symbf{a}_{\Omega}^{(3)}&=-\omega_3\symbf{a}_1^{(3)},\\
\symbf{a}_{\Omega}^{(4)}&=-\omega_4\symbf{a}_1^{(4)}.
\end{aligned}
\et{76}
\end{equation}

For $\symbf{p}_{\Omega}^{(n)}=\symbf{b}_{\Omega}^{(n)}$ and $\symbf{q}_1^{(n)}=\symbf{b}_1^{(n)}$, the central-moment-space collision form, after using the relationship between central-moment and raw-moment Hermite coefficients, is

\begin{equation}
\begin{aligned}
\symbf{a}_{\Omega}^{(2)}&=\symbf{b}_{\Omega}^{(2)}=-\omega_2\symbf{b}_1^{(2)},\\
\symbf{a}_{\Omega}^{(3)}-3\uv \symbf{a}_{\Omega}^{(2)}&=\symbf{b}_{\Omega}^{(3)}=-\omega_3\symbf{b}_1^{(3)},\\
\symbf{a}_{\Omega}^{(4)}-4\uv \symbf{a}_{\Omega}^{(3)}+6\uv\uv \symbf{a}_{\Omega}^{(2)}
&=\symbf{b}_{\Omega}^{(4)}=-\omega_4\symbf{b}_1^{(4)}.
\end{aligned}
\et{77}
\end{equation}

Equivalently,

\begin{align}
\symbf{a}_{\Omega}^{(2)}&=-\omega_2\symbf{a}_1^{(2)},\et{78}\\
\symbf{a}_{\Omega}^{(3)}&=-\omega_3\symbf{a}_1^{(3)}+3(\omega_3-\omega_2)\uv \symbf{a}_1^{(2)},\et{79}\\
\symbf{a}_{\Omega}^{(4)}&=-\omega_4\symbf{a}_1^{(4)}+4(\omega_4-\omega_3)\uv \symbf{a}_1^{(3)}\nonumber\\
&\quad-6(\omega_4+\omega_2-2\omega_3)\uv\uv \symbf{a}_1^{(2)}.\et{80}
\end{align}

For $\symbf{p}_{\Omega}^{(n)}=\symbf{d}_{\Omega}^{(n)}$ and $\symbf{q}_1^{(n)}=\symbf{d}_1^{(n)}$, the temperature-scaled central-moment collision form is

\begin{equation}
\begin{aligned}
\symbf{a}_{\Omega}^{(2)}&=(\sqrt{\theta})^{D+2}\symbf{d}_{\Omega}^{(2)}
=-\omega_2(\sqrt{\theta})^{D+2}\symbf{d}_1^{(2)},\\
\symbf{a}_{\Omega}^{(3)}-3\uv \symbf{a}_{\Omega}^{(2)}&=(\sqrt{\theta})^{D+3}\symbf{d}_{\Omega}^{(3)}\\
&=-\omega_3(\sqrt{\theta})^{D+3}\symbf{d}_1^{(3)},\\
\symbf{a}_{\Omega}^{(4)}-4\uv \symbf{a}_{\Omega}^{(3)}\\
&\quad+6\uv\uv \symbf{a}_{\Omega}^{(2)}-6(\theta-1)\delv \symbf{a}_{\Omega}^{(2)}
=(\sqrt{\theta})^{D+4}\symbf{d}_{\Omega}^{(4)}\\
&\quad=-\omega_4(\sqrt{\theta})^{D+4}\symbf{d}_1^{(4)}.
\end{aligned}
\et{81}
\end{equation}

After rearrangement, this becomes

\begin{align}
\symbf{a}_{\Omega}^{(2)}&=-\omega_2\symbf{a}_1^{(2)},\et{82}\\
\symbf{a}_{\Omega}^{(3)}&=-\omega_3\symbf{a}_1^{(3)}+3(\omega_3-\omega_2)\uv \symbf{a}_1^{(2)},\et{83}\\
\symbf{a}_{\Omega}^{(4)}&=-\omega_4\symbf{a}_1^{(4)}+4(\omega_4-\omega_3)\uv \symbf{a}_1^{(3)}\nonumber\\
&\quad-6(\omega_4+\omega_2-2\omega_3)\uv\uv \symbf{a}_1^{(2)}\nonumber\\
&\quad+6(\omega_4-\omega_2)(\theta-1)\delv \symbf{a}_1^{(2)}.\et{84}
\end{align}

\Es{82}{84} give the collision form proposed by the authors in 2019 \refc{56}. Comparing Eqs.~\eqref{eq:78}--\eqref{eq:80} with Eqs.~\eqref{eq:82}--\eqref{eq:84}, the second- and third-order terms are identical, whereas the fourth-order term in \E{84} contains an additional correction proportional to the temperature deviation $(\theta-1)$. Equation~\eqref{eq:75} shows that each order can formally be relaxed independently, regardless of whether raw-moment, central-moment or temperature-scaled central-moment Hermite coefficients are used.

The analysis in \secref{non-equilibrium}, however, shows that second-order moments are equivalent across these spaces. Provided that the discrete-velocity set has sufficient quadrature accuracy, the second moment of the velocity-discretized non-equilibrium distribution equals that of its continuous counterpart. Shan et al. \refc{45,71} showed that this requires at least 17 discrete velocities in two dimensions and 39 in three dimensions. With a sufficiently rich velocity set and an equilibrium expansion retained through at least third order, the viscous transport coefficients satisfy Galilean invariance exactly, irrespective of the space in which relaxation is performed \refc{55}. Cascaded \refc{17} and cumulant \refc{27} collision models on the D2Q9 or D3Q27 lattices cannot, by contrast, guarantee exact Galilean invariance of the viscous transport coefficients.

The macroscopic velocity is dimensional and changes under translation between inertial coordinate systems. In temperature-scaled central-moment space, each moment is independent of the macroscopic velocity and is nondimensionalized by the local speed of sound. As Eqs.~\eqref{eq:82}--\eqref{eq:84} demonstrate, the quantities relaxed in this space are genuine non-equilibrium quantities: the conserved contributions associated with momentum and internal energy are removed from the raw-moment relaxation targets, so moments of different orders remain independent. Taken together, Eqs.~\eqref{eq:76} and \eqref{eq:82}--\eqref{eq:84} show that relaxing each raw-moment non-equilibrium Hermite coefficient introduces coupling between third- and higher-order moments and lower-order moments, commonly termed crosstalk. Thus, moments of different orders are not truly relaxed independently in raw-moment space. The relaxation targets in a collision operator should therefore be invariant under translations of the coordinate system. Recent work by Li et al. \refc{77} further discusses the requirement of rotational invariance; a detailed treatment lies beyond the scope of this review.

\section{Regularization of the collision operator}\label{sec:regularization}

\subsection{Spatiotemporal discretization of the lattice Boltzmann equation}\label{sec:discretization}

The lattice Boltzmann equation discretized in velocity space, including a force term, is

\begin{equation}
\frac{\partial f_i}{\partial t}+\xv_i\!\cdot\!\nabla f_i=\Omega_i+F_i.
\et{85}
\end{equation}

Integrating \E{85} along a characteristic line and evaluating the right-hand-side integral with the trapezoidal rule gives

\begin{equation}
\begin{aligned}
f_i(\boldsymbol{x}+\xv_i,t+1)-f_i(\boldsymbol{x},t)\\
&=\frac12\big[\Omega_i(\boldsymbol{x}+\xv_i,t+1)\\
&\quad+\Omega_i(\boldsymbol{x},t)+F_i(\boldsymbol{x}+\xv_i,t+1)\\
&\quad+F_i(\boldsymbol{x},t)\big].
\end{aligned}
\et{86}
\end{equation}

For simplicity, the time step is set to $\Delta t=1$. Equation~\eqref{eq:86} is implicit because its right-hand side contains quantities at the next time step. To remove this implicitness, we introduce the transformed variable

\begin{equation}
\fbar_i=f_i-\frac12\Omega_i-\frac12F_i.
\et{87}
\end{equation}

Substitution of the new variable into \E{86} gives the explicit equation, in which all right-hand-side terms are evaluated at the current time step:

\begin{equation}
\fbar_i(\boldsymbol{x}+\xv_i,t+1)=\fbar_{p,i}(\boldsymbol{x},t)
=\fbar_i(\boldsymbol{x},t)+\Omega_i(\boldsymbol{x},t)+F_i(\boldsymbol{x},t).
\et{88}
\end{equation}

Here, $\fbar_{p,i}(\boldsymbol{x},t)$ denotes the particle distribution after collision and before streaming. Expanding the right-hand side of \E{88} in a Hermite basis and multiplying by the $n$th-order Hermite basis gives the Hermite coefficients of the right-hand side, following the operation in \E{6}. This operation can be viewed as a transformation from particle-distribution-function space to the corresponding Hermite-moment space. The general form is

\begin{equation}
\symbf{\qbar}_p^{(n)}=\symbf{\qbar}^{(n)}-\frac{1}{\tau_n}\left(\symbf{q}^{(n)}-\symbf{q}_0^{(n)}\right)+\symbf{q}_F^{(n)}.
\et{89}
\end{equation}

Equation~\eqref{eq:89} applies to raw, central and temperature-scaled central moments. Similarly, the transformation of the new variable into Hermite-moment space is

\begin{equation}
\symbf{\qbar}^{(n)}=\symbf{q}^{(n)}+\frac{1}{2\tau_n}\left(\symbf{q}^{(n)}-\symbf{q}_0^{(n)}\right)-\frac12\symbf{q}_F^{(n)}.
\et{90}
\end{equation}

The original variable can therefore be expressed in terms of the transformed variable as

\begin{equation}
\symbf{q}^{(n)}=\frac{2\tau_n}{2\tau_n+1}\symbf{\qbar}^{(n)}
+\frac{1}{2\tau_n+1}\symbf{q}_0^{(n)}+\frac{\tau_n}{2\tau_n+1}\symbf{q}_F^{(n)}.
\et{91}
\end{equation}

Substituting \E{91} into \E{89} gives the moment-space relaxation equation for the transformed distribution:

\begin{equation}
\begin{aligned}
\symbf{\qbar}_p^{(n)}&=\symbf{\qbar}^{(n)}-\frac{1}{\lambda_n}\left(\symbf{\qbar}^{(n)}-\symbf{q}_0^{(n)}\right)\\
&\quad+\left(1-\frac{1}{2\lambda_n}\right)\symbf{q}_F^{(n)},\qquad
\lambda_n=\tau_n+\frac12.
\end{aligned}
\et{92}
\end{equation}

The non-equilibrium moment of the transformed distribution is

\begin{equation}
\symbf{\qbar}_1^{(n)}=\symbf{\qbar}^{(n)}-\symbf{q}_0^{(n)}.
\et{93}
\end{equation}

The moment-space relaxation equation can thus be written as

\begin{equation}
\symbf{\qbar}_p^{(n)}=\symbf{q}_0^{(n)}+\left(1-\frac{1}{\lambda_n}\right)\symbf{\qbar}_1^{(n)}
+\left(1-\frac{1}{2\lambda_n}\right)\symbf{q}_F^{(n)},\qquad n\geq1.
\et{94}
\end{equation}

In raw-moment, central-moment and temperature-scaled central-moment spaces, respectively,

\begin{equation}
\begin{aligned}
\symbf{\abar}_p^{(n)}&=\symbf{a}_0^{(n)}+\left(1-\frac{1}{\lambda_n}\right)\symbf{\abar}_1^{(n)}
+\left(1-\frac{1}{2\lambda_n}\right)\symbf{a}_F^{(n)},\\
\symbf{\bbar}_p^{(n)}&=\symbf{b}_0^{(n)}+\left(1-\frac{1}{\lambda_n}\right)\symbf{\bbar}_1^{(n)}
+\left(1-\frac{1}{2\lambda_n}\right)\symbf{b}_F^{(n)},\\
\symbf{\dbar}_p^{(n)}&=\symbf{d}_0^{(n)}+\left(1-\frac{1}{\lambda_n}\right)\symbf{\dbar}_1^{(n)}
+\left(1-\frac{1}{2\lambda_n}\right)\symbf{d}_F^{(n)},\qquad n\geq1.
\end{aligned}
\et{95}
\end{equation}

When no force term is present, the collision equation starts from the second order because mass and momentum are conserved. For the transformed distribution, however, the first-order non-equilibrium moment is nonzero when a force is present. This point requires attention in multiphase models \refc{62--63} and immersed-boundary methods \refc{11}. We first discuss the relationship between the macroscopic quantities calculated from the transformed particle distribution and the actual physical quantities.

The zeroth-order moment is

\begin{equation}
\sum_i\fbar_i=\sum_i\left(f_i-\frac12\Omega_i-\frac12F_i\right)
=\sum_i f_i.
\et{96}
\end{equation}

Therefore, the physical density is

\begin{equation}
\rho=\bar\rho.
\et{97}
\end{equation}

The first-order moment is

\begin{equation}
\sum_i\fbar_i\xv_i=\sum_i\left(f_i-\frac12\Omega_i-\frac12F_i\right)\xv_i
=\sum_i\left(f_i-\frac12F_i\right)\xv_i.
\et{98}
\end{equation}

The physical velocity is therefore

\begin{equation}
\uv=\bar{\uv}+\frac{\Fv}{2\rho}.
\et{99}
\end{equation}

The trace of the second-order moment is

\begin{equation}
\begin{aligned}
\sum_i\fbar_i c_{i,\alpha}c_{i,\alpha}
&=\sum_i\left(f_i-\frac12\Omega_i-\frac12F_i\right)c_{i,\alpha}c_{i,\alpha}\\
&=\sum_i f_i c_{i,\alpha}c_{i,\alpha}.
\end{aligned}
\et{100}
\end{equation}

Here $\co_i=\xv_i-\uv$ is defined using the physical velocity in \E{99}. Defining $\bar\theta=(D\rho)^{-1}\sum_i\fbar_i|\xv_i-\uv|^2$, the physical temperature is

\begin{equation}
\theta=\bar\theta.
\et{101}
\end{equation}

If the transformed distribution is instead centred on $\bar{\uv}$, its central second moment gives $\theta-|\gv|^2/(4D)$, so \E{101} requires the definition just stated.

The density, velocity and temperature used in the equilibrium distribution and force term are all the actual physical quantities, without overbars. The first-order moment of the transformed distribution is not zero:

\begin{equation}
\begin{aligned}
\symbf{\abar}_1^{(1)}&=\sum_i\fbar_i^{(1)}\xv_i
=\sum_i\left(f_i^{(1)}-\frac12\Omega_i-\frac12F_i\right)\xv_i\\
&=\sum_i-\frac12F_i\xv_i=-\frac12\Fv.
\end{aligned}
\et{102}
\end{equation}

Similarly, the first-order central-moment non-equilibrium coefficient is

\begin{equation}
\begin{aligned}
\symbf{\bbar}_1^{(1)}&=\sum_i\fbar_i^{(1)}\co_i
=\sum_i\left(f_i^{(1)}-\frac12\Omega_i-\frac12F_i\right)\co_i\\
&=\sum_i-\frac12F_i\co_i=-\frac12\Fv.
\end{aligned}
\et{103}
\end{equation}

In temperature-scaled central-moment space,

\begin{equation}
\begin{aligned}
\symbf{\dbar}_1^{(1)}&=\frac{1}{(\sqrt{\theta})^D}\sum_i\fbar_i^{(1)}\boldsymbol{\eta}_i\\
&=\frac{1}{(\sqrt{\theta})^D}\sum_i\left(f_i^{(1)}-\frac12\Omega_i-\frac12F_i\right)\boldsymbol{\eta}_i\\
&=-\frac12\frac{\Fv}{(\sqrt{\theta})^{D+1}}.
\end{aligned}
\et{104}
\end{equation}

For collision in raw-moment space, the equilibrium part, non-equilibrium part and force term can all be used directly in their truncated Hermite forms to reconstruct the particle distribution. For collision in central-moment and temperature-scaled central-moment spaces, the discrete velocities depend on the local velocity and temperature. We therefore transform the collision result into raw-moment space before reconstructing the particle distribution. According to the collision space and truncation order, existing regularized models can now be analysed and discussed systematically. For convenience, the overbars on non-equilibrium moments and distributions are omitted in \secref{analysis}. In that subsection $\omega_n=1/\lambda_n$ is the discrete relaxation rate, rather than the continuous rate $1/\tau_n$ used in \secref{collision}.

\subsection{Analysis and discussion of regularized collision models}\label{sec:analysis}

\subsubsection{Ladd and Latt regularized models}\label{sec:ladd-latt}

Ladd et al. proposed the earliest forms of the regularized collision model in 1994 \refc{39} and 2001 \refc{41}. In their collision model, the equilibrium distribution was expanded to second order, and the non-equilibrium distribution was also retained only through second order. The post-collision distribution is

\begin{equation}
f_{p,i}^{\mathrm{reg}}=\omega_i\rho\left[1+\uv\!\cdot\!\xv_i
+\frac12\left(\uv\uv+\frac{\boldsymbol{\Pi}_p^{\mathrm{neq}}}{\rho}\right)
:\left(\xv_i\xv_i-\boldsymbol{I}\right)\right].
\et{105}
\end{equation}

All dimensional velocity quantities in this paper have been scaled by the speed of sound, so $c_s=1$ in \E{105}; a dimensional form would also require $c_s^{-2}$ and $c_s^{-4}$ in the linear and quadratic terms. The second-order non-equilibrium relaxation is

\begin{equation}
\boldsymbol{\Pi}_p^{\mathrm{neq}}=\left(1-\frac{1}{\lambda_{2,s}}\right)\overline{\boldsymbol{\Pi}}^{\mathrm{neq}}
+\left(1-\frac{1}{\lambda_{2,b}}\right)\frac{\Tr(\boldsymbol{\Pi}^{\mathrm{neq}})}{D}\boldsymbol{I}.
\et{106}
\end{equation}

Here, $\overline{\boldsymbol{\Pi}}^{\mathrm{neq}}$ is the trace-free part of the second-order stress tensor, $\Tr(\boldsymbol{\Pi}^{\mathrm{neq}})$ is its trace, $\lambda_{2,s}$ is the shear-viscosity relaxation time, and $\lambda_{2,b}$ is the bulk-viscosity relaxation time. The second-order stress tensor is obtained from the second moment of the non-equilibrium particle distribution:

\begin{equation}
\Pi_{\alpha\beta}^{\mathrm{neq}}=\sum_i f_i^{(1)}\xi_{i,\alpha}\xi_{i,\beta}.
\et{107}
\end{equation}

If the discrete velocities are those of standard lattices such as D2Q9, D3Q15, D3Q19 or D3Q27, the second-order stress tensor obtained from \E{107} has a nonzero trace, whereas the continuous second velocity moment of the non-equilibrium particle distribution is trace-free. In an exact calculation, there is therefore no bulk-viscosity term. The trace component appears because the discrete velocity set is not sufficiently large. Bulk viscosity usually arises from the excitation of internal degrees of freedom in polyatomic gases or from dense-fluid effects. The Ladd model is isothermal. Considering energy conservation, the eigenvalue of the bulk-viscosity mode, $1/\lambda_{2,b}$, should be zero. Because energy conservation is not enforced in the numerical implementation, this relaxation factor remains freely adjustable.

Latt et al. \refc{42} also proposed a regularized model in 2005. Their model did not separate the trace-free and trace parts of the second-order stress tensor and was simply

\begin{equation}
\boldsymbol{\Pi}_p^{\mathrm{neq}}=\left(1-\frac{1}{\lambda_2}\right)\boldsymbol{\Pi}^{\mathrm{neq}}.
\et{108}
\end{equation}

The Latt model was proposed after the Ladd model and can be viewed as the special case in which the shear- and bulk-viscosity relaxation factors are equal. It is therefore better interpreted as a particular formulation of the Ladd model than as a distinct collision model. The Ladd model is a regularized collision model in raw-moment space. Non-equilibrium moments of third and higher order, which lie beyond the isothermal Navier--Stokes equations, can be assigned unit relaxation factors and consequently make no contribution to the post-collision distribution.

\subsubsection{Zhang--Shan--Chen regularized model}\label{sec:zsc}

Zhang, Shan and Chen \refc{44} proposed a high-order regularized lattice Boltzmann model in 2006 based on a Hermite expansion and used it to simulate high-Knudsen-number non-continuum flows. They argued that both the equilibrium and non-equilibrium distributions should be projected onto the finite-dimensional Hilbert space spanned by the retained Hermite basis. Their formulation also included a force term. As discussed in \secref{discretization}, the first-order non-equilibrium Hermite coefficient is nonzero in the presence of an external force. The regularized post-collision distribution should therefore be written as

\begin{equation}
f_{p,i}^{\mathrm{reg}}=f_i^{(0)}+\omega_i\sum_{n=1}^{3}\frac{1}{n!}
\left(\symbf{a}_1^{(n)}+\symbf{a}_{\Omega}^{(n)}\right):\mathcal{H}^{(n)}(\xv_i)+F_i.
\et{109}
\end{equation}

Equivalently, the expression can be written in terms of the equilibrium, non-equilibrium and force Hermite coefficients as

\begin{equation}
\begin{aligned}
f_{p,i}^{\mathrm{reg}}={}&\omega_i\sum_{n=0}^{3}\frac{1}{n!}\symbf{a}_0^{(n)}:\mathcal{H}^{(n)}(\xv_i)\\
&+\omega_i\sum_{n=1}^{3}\left(1-\frac{1}{\lambda_n}\right)\frac{1}{n!}
\symbf{a}_1^{(n)}:\mathcal{H}^{(n)}(\xv_i)\\
&+\omega_i\sum_{n=1}^{3}\left(1-\frac{1}{2\lambda_n}\right)\frac{1}{n!}
\symbf{a}_F^{(n)}:\mathcal{H}^{(n)}(\xv_i).
\end{aligned}
\et{110}
\end{equation}

If all non-equilibrium moments of order $n\geq4$ are truncated, this is equivalent to setting all corresponding higher-order relaxation factors to one. The first-order non-equilibrium and force Hermite coefficients in \E{110} give

\begin{equation}
\left(1-\frac{1}{\lambda_1}\right)\left(-\frac{\Fv}{2}\right)
+\left(1-\frac{1}{2\lambda_1}\right)\Fv=\frac{\Fv}{2}.
\et{111}
\end{equation}

Thus, the first-order relaxation factor has no practical significance. In the original Zhang--Shan--Chen model, however, the reconstruction in \E{109} starts at second order, and the force term lacks the factor $1-1/(2\lambda_n)$. The original formulation is therefore not fully self-consistent when a force term is present.

\subsubsection{Mattila--Philippi--Hegele regularized model}\label{sec:mattila}

Mattila et al. \refc{49} proposed a high-order regularized lattice Boltzmann model based on an expansion in central-moment space for compressible thermal flows. We analyse it using the mathematical relationships between the central- and raw-moment expansions of the non-equilibrium coefficients in \secref{non-equilibrium} and the collision term in \secref{collision}. This model does not contain a force term, which is consequently omitted here. For convenience, the non-equilibrium Hermite coefficients and collision term are combined as in \E{95}; introducing the new collision form into \E{77} gives

\begin{equation}
\symbf{a}_{\Omega,s}^{(2)}=\symbf{b}_{\Omega,s}^{(2)}=(1-\omega_2)\symbf{b}_1^{(2)}.
\et{112}
\end{equation}

\begin{equation}
\symbf{a}_{\Omega,s}^{(3)}-3\uv \symbf{a}_{\Omega,s}^{(2)}
=\symbf{b}_{\Omega,s}^{(3)}=(1-\omega_3)\symbf{b}_1^{(3)}.
\et{113}
\end{equation}

\begin{equation}
\symbf{a}_{\Omega,s}^{(4)}-4\uv \symbf{a}_{\Omega,s}^{(3)}+6\uv\uv \symbf{a}_{\Omega,s}^{(2)}
=\symbf{b}_{\Omega,s}^{(4)}=(1-\omega_4)\symbf{b}_1^{(4)}=0.
\et{114}
\end{equation}

Here, the relaxation factors of fourth- and higher-order non-equilibrium moments are set to one for compressible thermal flows. In other words, higher-order kinetic central moments beyond the second-order stress tensor and third-order heat-flux vector are assumed to relax rapidly, so that these higher-order central-moment components of the post-collision distribution take their equilibrium values. Using the relationship between central- and raw-moment Hermite coefficients, the above equations can be rearranged as

\begin{equation}
\symbf{a}_{\Omega,s}^{(2)}=\symbf{a}_1^{(2)}-\omega_2\symbf{a}_1^{(2)}=\symbf{a}_1^{(2)}+\symbf{a}_{\Omega}^{(2)}.
\et{115}
\end{equation}

\begin{equation}
\begin{aligned}
\symbf{a}_{\Omega,s}^{(3)}={}&\symbf{a}_1^{(3)}+\left[-\omega_3\symbf{a}_1^{(3)}
+(\omega_3-\omega_2)3\uv \symbf{a}_1^{(2)}\right]\\
={}&\symbf{a}_1^{(3)}+\symbf{a}_{\Omega}^{(3)}.
\end{aligned}
\et{116}
\end{equation}

\begin{equation}
\begin{aligned}
\symbf{a}_{\Omega,s}^{(4)}={}&4\uv \symbf{a}_{\Omega,s}^{(3)}-6\uv\uv \symbf{a}_{\Omega,s}^{(2)}\\
={}&4\uv \symbf{a}_1^{(3)}\\
&\quad-6\uv\uv \symbf{a}_1^{(2)}\\
&+4\uv(-\omega_3\symbf{a}_1^{(3)}
+(\omega_3-\omega_2)3\uv \symbf{a}_1^{(2)})\\
&-6\uv\uv(-\omega_2\symbf{a}_1^{(2)})\\
={}&4\uv \symbf{a}_1^{(3)}-6\uv\uv \symbf{a}_1^{(2)}+4\uv \symbf{a}_{\Omega}^{(3)}\\
&\quad-6\uv\uv \symbf{a}_{\Omega}^{(2)}.
\end{aligned}
\et{117}
\end{equation}

Equations~\eqref{eq:115} and \eqref{eq:116} show that the second- and third-order terms in the new collision form are the sums of the original collision terms and the corresponding non-equilibrium terms. According to \E{44}, after the fourth-order non-equilibrium central moment $\symbf{b}_1^{(4)}$ is removed, the fourth-order raw-moment coefficient is

\begin{equation*}
\symbf{a}_1^{(4)}=4\uv \symbf{a}_1^{(3)}-6\uv\uv \symbf{a}_1^{(2)},
\qquad \symbf{b}_1^{(4)}=0.
\end{equation*}

This is exactly the non-relaxed contribution in \E{117}. Substituting the filtered expression for $\symbf{a}_1^{(4)}$ into \E{80} gives the form of \E{117}. Thus, within the theoretical framework developed here, the regularization process is equivalent to setting the central-moment relaxation factors of fourth and higher order in \E{114} to one, allowing the higher-order kinetic moments to relax rapidly to equilibrium.

For isothermal flows, the relaxation factors of third- and higher-order central moments can all be set to one, giving

\begin{equation}
\begin{aligned}
\symbf{a}_{\Omega,s}^{(2)}&=\symbf{b}_{\Omega,s}^{(2)}=(1-\omega_2)\symbf{b}_1^{(2)},\\
\symbf{a}_{\Omega,s}^{(3)}-3\uv \symbf{a}_{\Omega,s}^{(2)}&=\symbf{b}_{\Omega,s}^{(3)}=(1-\omega_3)\symbf{b}_1^{(3)}=0,\\
\symbf{a}_{\Omega,s}^{(4)}-4\uv \symbf{a}_{\Omega,s}^{(3)}+6\uv\uv \symbf{a}_{\Omega,s}^{(2)}&=\symbf{b}_{\Omega,s}^{(4)}=(1-\omega_4)\symbf{b}_1^{(4)}=0.
\end{aligned}
\et{118}
\end{equation}

Algebraic simplification yields

\begin{equation}
\symbf{a}_{\Omega,s}^{(2)}=\symbf{a}_1^{(2)}-\omega_2\symbf{a}_1^{(2)}=\symbf{a}_1^{(2)}+\symbf{a}_{\Omega}^{(2)}.
\et{119}
\end{equation}

\begin{equation}
\symbf{a}_{\Omega,s}^{(3)}=3\uv \symbf{a}_{\Omega,s}^{(2)}
=3\uv \symbf{a}_1^{(2)}-\omega_2\left(3\uv \symbf{a}_1^{(2)}\right).
\et{120}
\end{equation}

\begin{equation}
\begin{aligned}
\symbf{a}_{\Omega,s}^{(4)}&=4\uv \symbf{a}_{\Omega,s}^{(3)}-6\uv\uv \symbf{a}_{\Omega,s}^{(2)}\\
&=6\uv\uv \symbf{a}_{\Omega,s}^{(2)}\\
&=6\uv\uv \symbf{a}_1^{(2)}+6\uv\uv\left(-\omega_2\symbf{a}_1^{(2)}\right).
\end{aligned}
\et{121}
\end{equation}

Mattila et al. \refc{49} noted that, for compressible thermal flows, the equilibrium distribution should retain terms through fourth order and the non-equilibrium distribution should also be retained through fourth order. However, the fourth-order central moment in the non-equilibrium part is dissipative and should not depend on the fifth-order equilibrium moment; it must therefore be filtered, leaving only the convective contribution. Denote this filtered raw-moment coefficient by $\symbf{b}_{1,a}^{(4)}$, as in the Chinese source; the additional subscript $a$ distinguishes it from the central moment $\symbf{b}_1^{(4)}$. Written with explicit indices,

\begin{equation}
\begin{aligned}
b_{1,a;\alpha\beta\gamma\delta}^{(4)}={}&[4\uv \symbf{a}_1^{(3)}-6\uv\uv \symbf{a}_1^{(2)}]_{\alpha\beta\gamma\delta}\\
={}&(u_\alpha a_{1,\beta\gamma\delta}^{(3)}
+u_\beta a_{1,\alpha\gamma\delta}^{(3)})\\
&+(u_\gamma a_{1,\alpha\beta\delta}^{(3)}
+u_\delta a_{1,\alpha\beta\gamma}^{(3)})\\
&-(u_\alpha u_\beta a_{1,\gamma\delta}^{(2)}
+u_\alpha u_\gamma a_{1,\beta\delta}^{(2)})\\
&-(u_\alpha u_\delta a_{1,\beta\gamma}^{(2)}
+u_\beta u_\gamma a_{1,\alpha\delta}^{(2)})\\
&-(u_\beta u_\delta a_{1,\alpha\gamma}^{(2)}
+u_\gamma u_\delta a_{1,\alpha\beta}^{(2)}).
\end{aligned}
\et{122}
\end{equation}

Mattila et al.'s formulation is mathematically equivalent to \E{114}: regularization is performed in central-moment space by setting the relaxation factors of fourth- and higher-order moments to one, or, equivalently, by filtering fourth- and higher-order non-equilibrium central moments. In the Zhang--Shan--Chen model, filtering in raw-moment space removes the entire fourth-order non-equilibrium contribution. In central-moment space, only $\symbf{b}_1^{(4)}=\symbf{a}_1^{(4)}-4\uv\symbf{a}_1^{(3)}+6\uv\uv\symbf{a}_1^{(2)}$ is removed; the remaining convective contribution is reconstructed from lower-order moments and velocity.

In the isothermal regime, the third- and higher-order non-equilibrium central moments are all filtered. The resulting raw-moment reconstructions are:

\begin{equation}
\begin{aligned}
b_{1,a;\alpha\beta\gamma}^{(3)}={}&a_{1,\alpha\beta\gamma}^{(3)}
-b_{1,\alpha\beta\gamma}^{(3)}\\={}&[3\uv \symbf{a}_1^{(2)}]_{\alpha\beta\gamma}\\
={}&u_\alpha a_{1,\beta\gamma}^{(2)}+u_\beta a_{1,\alpha\gamma}^{(2)}
+u_\gamma a_{1,\alpha\beta}^{(2)}.
\end{aligned}
\et{123}
\end{equation}

\begin{equation}
\begin{aligned}
b_{1,a;\alpha\beta\gamma\delta}^{(4)}={}&[6\uv\uv \symbf{a}_1^{(2)}]_{\alpha\beta\gamma\delta}\\
={}&u_\alpha u_\beta a_{1,\gamma\delta}^{(2)}+u_\alpha u_\gamma a_{1,\beta\delta}^{(2)}
+u_\alpha u_\delta a_{1,\beta\gamma}^{(2)}\\
&+u_\beta u_\gamma a_{1,\alpha\delta}^{(2)}+u_\beta u_\delta a_{1,\alpha\gamma}^{(2)}
+u_\gamma u_\delta a_{1,\alpha\beta}^{(2)}.
\end{aligned}
\et{124}
\end{equation}

The first terms on the right-hand sides of \E{120} and \E{121} are equivalent to the expressions in \E{123} and \E{124}, respectively, whereas the second terms represent their relaxation. It is worth noting that Mattila's single-relaxation collision form can treat only flows with Prandtl number equal to one. Even when a multiple-relaxation form is adopted,
the Galilean invariance of the thermal conductivity cannot be guaranteed. This was demonstrated numerically by Chen et al. \refc{46} and Shan et al. \refc{55}; \E{116} shows the correction $(\omega_3-\omega_2)3\uv \symbf{a}_1^{(2)}$ that must be included in the raw-moment collision to cancel this spurious coupling.

\subsubsection{Malaspinas regularized model}\label{sec:malaspinas}

Malaspinas \refc{48} proposed an isothermal recursive regularized model in 2015. The recursion consists of a recursive construction of the equilibrium Hermite coefficients and a recursive construction of the non-equilibrium Hermite coefficients. When only the isothermal Navier--Stokes equations are considered, the recursive relationship for the equilibrium Hermite coefficients is simple and follows by mathematical induction:

\begin{equation}
\symbf{a}_0^{(n)}=\symbf{a}_0^{(n-1)}\uv,\qquad \symbf{a}_0^{(0)}=\rho.
\et{125}
\end{equation}

The recursive relationship for the non-equilibrium Hermite coefficients is derived from the recursion for the equilibrium coefficients, the Euler equations obtained from the zeroth-order CE expansion, and the relationship between the non-equilibrium and equilibrium Hermite coefficients and their space-time derivatives obtained from the first-order CE expansion. It is

\begin{equation}
\begin{aligned}
a_{1,\alpha_1\cdots\alpha_n}^{(n)}={}&
a_{1,\alpha_1\cdots\alpha_{n-1}}^{(n-1)}u_{\alpha_n}\\
&+[u_{\alpha_1}\cdots u_{\alpha_{n-2}}
a_{1,\alpha_{n-1}\alpha_n}^{(2)}\\
&\quad+\perm(\alpha_n)],\qquad n\geq3.
\end{aligned}
\et{126}
\end{equation}

Here, $\perm(\alpha_n)$ denotes all permutations of the indices other than $\alpha_n$. Equation~\eqref{eq:126} shows that the $n$th-order non-equilibrium Hermite coefficient can be obtained recursively from the $(n-1)$th-order coefficient, the second-order non-equilibrium Hermite coefficient and the local velocity.

\noindent\begin{minipage}{\columnwidth}
\setlength{\parindent}{1em}
For two-dimensional problems, Malaspinas started from the second-order Taylor expansion of the Maxwell--Boltzmann equilibrium distribution and retained selected terms of orders three and four, whose coefficients were determined from the recursive relationship. For the D2Q9 lattice, following the ideas of D'Humières \refc{15--16} and Dellar \refc{78}, the number of basis functions is taken to be equal to the number of discrete velocities. In addition to the six basis functions from zeroth to second order, the expansion contains three basis functions of third and fourth order:
\end{minipage}

\begin{equation}
\begin{aligned}
f_i^{(0)}=\omega_i\rho\bigg[&1+\xv_i\!\cdot\!\uv
+\frac12\mathcal{H}_i^{(2)}:\uv\uv\\
&+\frac12\left(\mathcal{H}_{i,xxy}^{(3)}u_x^2u_y
+\mathcal{H}_{i,xyy}^{(3)}u_y^2u_x\right)\\
&+\frac14\mathcal{H}_{i,xxyy}^{(4)}u_x^2u_y^2\bigg].
\end{aligned}
\et{127}
\end{equation}

The corresponding non-equilibrium expansion is given in \E{128}.

\begin{equation}
\begin{aligned}
f_i^{(1)}=\omega_i\bigg[&\frac12\mathcal{H}_i^{(2)}:\symbf{a}_1^{(2)}\\
&+\frac12\left(\mathcal{H}_{i,xxy}^{(3)}a_{1,xxy}^{(3)}
+\mathcal{H}_{i,xyy}^{(3)}a_{1,xyy}^{(3)}\right)\\
&+\frac14\mathcal{H}_{i,xxyy}^{(4)}a_{1,xxyy}^{(4)}\bigg].
\end{aligned}
\et{128}
\end{equation}

The expansion bases for the equilibrium and non-equilibrium distributions correspond to one another. The second-order non-equilibrium Hermite coefficient is obtained from the second moment of the non-equilibrium particle distribution, whereas the third- and fourth-order coefficients are obtained recursively. For three-dimensional problems on the D3Q27 lattice, the equilibrium expansion is identical to the two-dimensional one through second order, but is continued through sixth order, giving 27 Hermite basis functions, as shown in \E{129}.

\begin{equation}
\begin{aligned}
f_i^{(0)}=\omega_i\rho\bigg\{&1+\xv_i\!\cdot\!\uv
+\tfrac12\mathcal{H}_i^{(2)}:\uv\uv\\
&+\tfrac12\bigl[\mathcal{H}_{i,xxy}^{(3)}u_x^2u_y
+\mathcal{H}_{i,xxz}^{(3)}u_x^2u_z\\
&\qquad+\mathcal{H}_{i,xyy}^{(3)}u_xu_y^2
+\mathcal{H}_{i,xzz}^{(3)}u_xu_z^2\\
&\qquad+\mathcal{H}_{i,yzz}^{(3)}u_yu_z^2
+\mathcal{H}_{i,yyz}^{(3)}u_y^2u_z\\
&\qquad+2\mathcal{H}_{i,xyz}^{(3)}u_xu_yu_z\bigr]\\
&+\tfrac14\bigl[\mathcal{H}_{i,xxyy}^{(4)}u_x^2u_y^2
+\mathcal{H}_{i,xxzz}^{(4)}u_x^2u_z^2\\
&\qquad+\mathcal{H}_{i,yyzz}^{(4)}u_y^2u_z^2
+2\mathcal{H}_{i,xyzz}^{(4)}u_xu_yu_z^2\\
&\qquad+2\mathcal{H}_{i,xyyz}^{(4)}u_xu_y^2u_z
+2\mathcal{H}_{i,xxyz}^{(4)}u_x^2u_yu_z\bigr]\\
&+\tfrac14\bigl[\mathcal{H}_{i,xxyzz}^{(5)}u_x^2u_yu_z^2
+\mathcal{H}_{i,xxyyz}^{(5)}u_x^2u_y^2u_z\\
&\qquad+\mathcal{H}_{i,xyyzz}^{(5)}u_xu_y^2u_z^2\bigr]\\
&+\tfrac18\mathcal{H}_{i,xxyyzz}^{(6)}u_x^2u_y^2u_z^2\bigg\}.
\end{aligned}
\et{129}
\end{equation}

The non-equilibrium particle distribution uses the same Hermite basis. Its second-order coefficient is evaluated by projection, and its higher-order coefficients are obtained recursively:
\begin{equation}
\begin{aligned}
f_i^{(1)}=\omega_i\bigg\{&\tfrac12\mathcal{H}_i^{(2)}:\symbf{a}_1^{(2)}\\
&+\tfrac12\bigl[\mathcal{H}_{i,xxy}^{(3)}a_{1,xxy}^{(3)}
+\mathcal{H}_{i,xxz}^{(3)}a_{1,xxz}^{(3)}\\
&\qquad+\mathcal{H}_{i,xyy}^{(3)}a_{1,xyy}^{(3)}
+\mathcal{H}_{i,xzz}^{(3)}a_{1,xzz}^{(3)}\\
&\qquad+\mathcal{H}_{i,yzz}^{(3)}a_{1,yzz}^{(3)}
+\mathcal{H}_{i,yyz}^{(3)}a_{1,yyz}^{(3)}\\
&\qquad+2\mathcal{H}_{i,xyz}^{(3)}a_{1,xyz}^{(3)}\bigr]\\
&+\tfrac14\bigl[\mathcal{H}_{i,xxyy}^{(4)}a_{1,xxyy}^{(4)}
+\mathcal{H}_{i,xxzz}^{(4)}a_{1,xxzz}^{(4)}\\
&\qquad+\mathcal{H}_{i,yyzz}^{(4)}a_{1,yyzz}^{(4)}
+2\mathcal{H}_{i,xyzz}^{(4)}a_{1,xyzz}^{(4)}\\
&\qquad+2\mathcal{H}_{i,xyyz}^{(4)}a_{1,xyyz}^{(4)}
+2\mathcal{H}_{i,xxyz}^{(4)}a_{1,xxyz}^{(4)}\bigr]\\
&+\tfrac14\bigl[\mathcal{H}_{i,xxyzz}^{(5)}a_{1,xxyzz}^{(5)}
+\mathcal{H}_{i,xxyyz}^{(5)}a_{1,xxyyz}^{(5)}\\
&\qquad+\mathcal{H}_{i,xyyzz}^{(5)}a_{1,xyyzz}^{(5)}\bigr]\\
&+\tfrac18\mathcal{H}_{i,xxyyzz}^{(6)}a_{1,xxyyzz}^{(6)}\bigg\}.
\end{aligned}
\et{130}
\end{equation}

The final regularized collision and streaming scheme is

\begin{equation}
f_i(\boldsymbol{x}+\xv_i,t+1)=f_i^{(0)}+\left(1-\frac{1}{\lambda}\right)f_i^{(1)}.
\et{131}
\end{equation}

In this model, the recursively obtained third- and higher-order non-equilibrium kinetic moments relax with the same factor as the second-order stress tensor. This differs in the higher-order terms from the Ladd \refc{39} and Latt \refc{42} models in the construction of the equilibrium distribution and in the reconstruction and relaxation of the non-equilibrium distribution. The bases used to reconstruct the equilibrium and non-equilibrium particle distributions are identical to the bases generated by one-dimensional tensor products in the cascaded collision model \refc{19}.

In the isothermal regime, the higher-order non-equilibrium moments obtained from the recursive relationship for the non-equilibrium Hermite coefficients are equivalent to those in the regularized model of K. Mattila et al. \refc{49}, which is based on a central-moment expansion. For example, the third- and fourth-order expressions in \E{123} and \E{124} are identical to the coefficients obtained from the recursive expression in \E{126}. Higher-order terms can be obtained by expanding the relationship between raw- and central-moment non-equilibrium Hermite coefficients in \E{45} to higher order and filtering the third- and higher-order central-moment non-equilibrium Hermite coefficients. The resulting fourth- and higher-order raw-moment coefficients are consistent with the recursive expression in \E{126}. Thus, Malaspinas's recursive relationship for the non-equilibrium Hermite coefficients can also be obtained by a central-moment expansion followed by filtering of the non-equilibrium central-moment Hermite coefficients above second order.

According to the CE expansion in \secref{truncation}, the $k$th-order contribution to the second-order non-equilibrium Hermite coefficient depends on the $(k+2)$th-order equilibrium Hermite coefficient. For example, the highest-order terms in $\symbf{a}_1^{(2)}$, $\symbf{a}_2^{(2)}$ and $\symbf{a}_3^{(2)}$ depend on $\symbf{a}_0^{(3)}$, $\symbf{a}_0^{(4)}$ and $\symbf{a}_0^{(5)}$, respectively. Capturing all three of these contributions exactly would require the equilibrium Hermite expansion to be retained through at least fifth order. In simulations of the isothermal N--S equations, the equilibrium Hermite expansion is usually truncated at third order rather than continued to infinite order. Consequently, the second-order non-equilibrium Hermite coefficient obtained from the second moment of the non-equilibrium particle distribution still contains spurious modes caused by the insufficient discrete-velocity set.

In 2018, Jacob et al. \refc{51} further improved Malaspinas's recursive regularized model. They calculated the second-order non-equilibrium coefficient by finite differences and combined it with the estimate obtained from the second moment of the non-equilibrium particle distribution through a weighted average. Numerical simulations on the D3Q19 lattice led to a hybrid recursive regularized model, whose controllability was also investigated. This hybrid model introduces additional hyperviscosity and thereby enhances numerical stability. In 2019, Chen et al. \refc{79} proposed another regularized model and claimed that it was equivalent to Malaspinas's recursive regularized model; however, the mathematical equivalence has not yet been clarified.

{\titlespacing*{\subsubsection}{0pt}{3.25ex}{1.5ex}
\subsubsection{Li--Shi--Shan regularized model}\label{sec:li-shi-shan}}

Li et al. \refc{56} proposed, in 2019, a high-order regularized lattice Boltzmann model based on an expansion in temperature-scaled central-moment space for compressible thermal flows. We analyse it using the mathematical relationships between the collision term and the non-equilibrium coefficients expanded in temperature-scaled central-moment and raw-moment spaces in \secrefs{collision}{non-equilibrium}. To maintain consistency with the analysis in \secref{mattila}, the temperature-scaled central-moment non-equilibrium Hermite coefficients and their relaxation are combined. On the basis of \E{81}, define

\begin{align}
\symbf{a}_{\Omega,s}^{(2)}&=(\sqrt{\theta})^{D+2}\symbf{d}_{\Omega,s}^{(2)}\notag\\[-1pt]
&=(1-\omega_2)(\sqrt{\theta})^{D+2}\symbf{d}_1^{(2)},\et{132}\\[0pt]
\symbf{a}_{\Omega,s}^{(3)}-3\uv \symbf{a}_{\Omega,s}^{(2)}
&=(\sqrt{\theta})^{D+3}\symbf{d}_{\Omega,s}^{(3)}\notag\\[-1pt]
&=(1-\omega_3)(\sqrt{\theta})^{D+3}\symbf{d}_1^{(3)},\et{133}\\[0pt]
\symbf{a}_{\Omega,s}^{(4)}&-4\uv \symbf{a}_{\Omega,s}^{(3)}
  +6\uv\uv \symbf{a}_{\Omega,s}^{(2)}\notag\\[-1pt]
&\quad-6(\theta-1)\delv \symbf{a}_{\Omega,s}^{(2)}\notag\\[-1pt]
&\qquad=(\sqrt{\theta})^{D+4}\symbf{d}_{\Omega,s}^{(4)}\notag\\[-1pt]
&\qquad=(1-\omega_4)(\sqrt{\theta})^{D+4}\symbf{d}_1^{(4)}=0.\et{134}
\end{align}

Using the relationship between temperature-scaled central-moment and raw-moment non-equilibrium Hermite coefficients, these equations become

\begin{equation}
\begin{aligned}
\symbf{a}_{\Omega,s}^{(2)}={}&(1-\omega_2)\symbf{a}_1^{(2)},\\[1pt]
\symbf{a}_{\Omega,s}^{(3)}={}&\symbf{a}_1^{(3)}+[-\omega_3\symbf{a}_1^{(3)}\\
&\quad+(\omega_3-\omega_2)3\uv \symbf{a}_1^{(2)}]\\
={}&\symbf{a}_1^{(3)}+\symbf{a}_{\Omega}^{(3)}.
\\[2pt]
\symbf{a}_{\Omega,s}^{(4)}={}&4\uv \symbf{a}_{\Omega,s}^{(3)}-6\uv\uv \symbf{a}_{\Omega,s}^{(2)}\\
&\quad+6(\theta-1)\delv \symbf{a}_{\Omega,s}^{(2)}\\
={}&(4\uv \symbf{a}_1^{(3)}-6\mathcal{U}_{-}\symbf{a}_1^{(2)})\\
&+4\uv(-\omega_3\symbf{a}_1^{(3)}
+(\omega_3-\omega_2)3\uv \symbf{a}_1^{(2)})\\
&-6\mathcal{U}_{-}(-\omega_2\symbf{a}_1^{(2)})\\
={}&\{4\uv \symbf{a}_1^{(3)}-6[\uv\uv-(\theta-1)\delv]\symbf{a}_1^{(2)}\}\\
&\quad+4\uv \symbf{a}_{\Omega}^{(3)}-6\mathcal{U}_{-}\symbf{a}_{\Omega}^{(2)}.
\end{aligned}
\et{135}
\end{equation}

Here, $\mathcal{U}_{-}=\uv\uv-(\theta-1)\delv$. These expressions show that the second- and third-order collisions are obtained by combining the raw-moment non-equilibrium Hermite coefficients in the collision forms of \E{78} and \E{79}. In the fourth-order collision, the raw-moment coefficient is the form obtained after filtering the fourth-order temperature-scaled central-moment Hermite coefficient:

\begin{equation}
\symbf{a}_1^{(4)}=4\uv \symbf{a}_1^{(3)}-6\left[\uv\uv-(\theta-1)\delv\right]\symbf{a}_1^{(2)}.
\et{136}
\end{equation}

Substitution of this expression into \E{84}, followed by addition of the non-equilibrium coefficient, gives \E{134}. The final post-collision particle distribution is

\begin{equation}
f_{p,i}=f_i^{(0)}+\omega_i\sum_{n=2}^{4}\frac{1}{n!}
\symbf{a}_{\Omega,s}^{(n)}:\mathcal{H}^{(n)}(\xv_i).
\et{137}
\end{equation}

The collision form proposed by Li et al. based on a Hermite expansion in temperature-scaled central-moment space can therefore be interpreted as setting the relaxation factors of fourth- and higher-order non-equilibrium Hermite moments in that space to one, so that higher-order moments rapidly relax to equilibrium. The Zhang--Shan--Chen, Mattila--Philippi--Hegele and Li--Shi--Shan models perform this filtering in raw-moment, central-moment and temperature-scaled central-moment spaces, respectively. All three therefore set the relaxation factors of the corresponding fourth- and higher-order non-equilibrium moments to one. Their essential difference is the space in which collision is performed. The relationships and differences among these models become transparent in the present analytical framework. The force-inclusive form of the raw-moment, central moment and temperature-scaled central-moment regularized model are elaborated in the appendix.

{\titlespacing*{\subsubsection}{0pt}{3.25ex}{1.5ex}
\subsubsection{Coreixas et al. regularized model}\label{sec:coreixas}}

In 2017, Coreixas et al. \refc{50} extended Malaspinas's isothermal recursive regularized model to the compressible thermal regime and developed a high-order recursive regularized lattice Boltzmann model. Its equilibrium Hermite coefficients satisfy

\begin{equation}
\begin{aligned}
a_{0,\alpha_1\cdots\alpha_n}^{(n)}={}&u_{\alpha_n}
a_{0,\alpha_1\cdots\alpha_{n-1}}^{(n-1)}\\
&+(\theta-1)\sum_{i=1}^{n-1}\delta_{\alpha_i\alpha_n}
a_{0,\beta_i}^{(n-2)},\qquad n\geq2.
\end{aligned}
\et{138}
\end{equation}

Here $\beta_i$ is the index list $(\alpha_1,\ldots,\alpha_{n-1})$ with $\alpha_i$ omitted, and $\beta_{ij}$ is the same list with both $\alpha_i$ and $\alpha_j$ omitted. Thus, $\beta_i$ and $\beta_{ij}$ contain $n-2$ and $n-3$ indices, respectively; neither contains $\alpha_n$. Compared with Malaspinas's isothermal recursive regularized model, the equilibrium recursion contains an additional temperature-dependent term. The recursive relationship for the non-equilibrium Hermite coefficients is likewise derived from the equilibrium recursion and the zeroth- and first-order CE expansions, as given in \E{139}.

\begin{equation}
\begin{aligned}
a_{1,\alpha_1\cdots\alpha_n}^{(n)}={}&u_{\alpha_n}a_{1,\alpha_1\cdots\alpha_{n-1}}^{(n-1)}
+(\theta-1)\sum_{i=1}^{n-1}\delta_{\alpha_i\alpha_n}a_{1,\beta_i}^{(n-2)}\\
&+\frac{1}{\rho}\sum_{i=1}^{n-1}a_{0,\beta_i}^{(n-2)}a_{1,\alpha_i\alpha_n}^{(2)}\\
&+\frac{1}{\rho}\sum_{i=1}^{n-1}\sum_{j>i}^{n-1}a_{0,\beta_{ij}}^{(n-3)}\bigl(\\
&\qquad a_{1,\alpha_i\alpha_j\alpha_n}^{(3)}
-u_{\alpha_i}a_{1,\alpha_j\alpha_n}^{(2)}
-u_{\alpha_j}a_{1,\alpha_i\alpha_n}^{(2)}\\
&\qquad-u_{\alpha_n}a_{1,\alpha_i\alpha_j}^{(2)}\bigr),\\
&\hfill n\geq4.
\end{aligned}
\et{139}
\end{equation}

\noindent
The resulting recursion is algebraically involved. The fourth-order non-equilibrium Hermite coefficient can be arranged as shown in \E{140}.

\begin{equation}
\begin{aligned}
a_{1,\alpha\beta\gamma\delta}^{(4)}={}&u_\delta a_{1,\alpha\beta\gamma}^{(3)}\\
&+(\theta-1)\bigl(\delta_{\alpha\delta}a_{1,\beta\gamma}^{(2)}
+\delta_{\beta\delta}a_{1,\alpha\gamma}^{(2)}\\
&\qquad+\delta_{\gamma\delta}a_{1,\alpha\beta}^{(2)}\bigr)\\
&+\frac{1}{\rho}\bigl(a_{0,\beta\gamma}^{(2)}a_{1,\alpha\delta}^{(2)}
+a_{0,\alpha\gamma}^{(2)}a_{1,\beta\delta}^{(2)}\\
&\qquad+a_{0,\alpha\beta}^{(2)}a_{1,\gamma\delta}^{(2)}\bigr)\\
&+\frac{a_{0,\gamma}^{(1)}}{\rho}\bigl(a_{1,\alpha\beta\delta}^{(3)}
-u_\alpha a_{1,\beta\delta}^{(2)}\\
&\qquad-u_\beta a_{1,\alpha\delta}^{(2)}
-u_\delta a_{1,\alpha\beta}^{(2)}\bigr)\\
&+\frac{a_{0,\beta}^{(1)}}{\rho}\bigl(a_{1,\alpha\gamma\delta}^{(3)}
-u_\alpha a_{1,\gamma\delta}^{(2)}\\
&\qquad-u_\gamma a_{1,\alpha\delta}^{(2)}
-u_\delta a_{1,\alpha\gamma}^{(2)}\bigr)\\
&+\frac{a_{0,\alpha}^{(1)}}{\rho}\bigl(a_{1,\beta\gamma\delta}^{(3)}
-u_\beta a_{1,\gamma\delta}^{(2)}\\
&\qquad-u_\gamma a_{1,\beta\delta}^{(2)}
-u_\delta a_{1,\beta\gamma}^{(2)}\bigr)\\
={}&4\uv\symbf{a}_1^{(3)}
-6[\uv\uv-(\theta-1)\delv]\symbf{a}_1^{(2)}.
\end{aligned}
\et{140}
\end{equation}

Equation~\eqref{eq:140} is equivalent to the fourth-order non-equilibrium moment of the Li--Shi--Shan model derived in temperature-scaled central-moment space in \E{135}. Higher-order non-equilibrium Hermite coefficients can likewise be obtained by applying the corresponding filtering operation through transformations analogous to Eqs.~\eqref{eq:48} and \eqref{eq:49}. Two equivalences between non-equilibrium reconstructions emerge: the Mattila--Philippi--Hegele reconstruction corresponds to Malaspinas's recursion, and the Li--Shi--Shan reconstruction corresponds to that of Coreixas et al. These equivalences do not imply identical MRT collision operators. These equivalences show that, in the isothermal regime, the raw-moment recursion follows directly from the transformation between central- and raw-moment Hermite coefficients. In the compressible thermal regime, it follows analogously from the transformation between temperature-scaled central-moment and raw-moment Hermite coefficients.

After reconstructing the non-equilibrium Hermite coefficients, the collision term in the Coreixas et al. regularized model is

\begin{equation}
\symbf{a}_{\Omega}^{(n)}=-\frac{\symbf{a}_1^{(n)}}{\lambda_n}.
\et{141}
\end{equation}

The final reconstruction of the post-collision particle distribution is

\begin{equation}
f_{p,i}=f_i^{(0)}+\omega_i\sum_{n=2}^{4}\frac{1}{n!}
\left(1-\frac{1}{\lambda_n}\right)\symbf{a}_1^{(n)}:\mathcal{H}^{(n)}(\xv_i).
\et{142}
\end{equation}

Compared with the Li--Shi--Shan regularized collision model in Eqs.~\eqref{eq:132}--\eqref{eq:134}, the high-order recursive regularized model of Coreixas et al. cannot guarantee Galilean invariance of the thermal conductivity even when a multiple-relaxation model is used. The Li--Shi--Shan model provides recursive forms for both the non-equilibrium Hermite coefficients and the collision term. This property illustrates the mathematical and physical advantages of expansion and regularization in temperature-scaled central-moment space.

\section{Conclusions}\label{sec:conclusions}

This article has provided a systematic and rigorous theoretical review of regularized collision models for lattice Boltzmann methods. By expanding the equilibrium distribution, non-equilibrium distribution, force term and collision term in Hermite bases with different variables, we have situated each theoretical model within a unified framework and clarified the relationships among them. The principal findings are as follows.

\begin{enumerate}[label=\arabic*),wide=\parindent]
\item Every regularized collision model can be obtained within the theoretical framework of Hermite expansions.
\item In the recursive regularized models of Malaspinas and Coreixas, the non-equilibrium distribution can be obtained recursively to arbitrarily high order. For kinetic moments beyond the moments considered here---the second-order stress tensor and the third-order heat-flux vector---the relaxation time need not be one. In the former model, all such moments relax with the relaxation time related to the shear viscosity, whereas in the latter model each order of the raw-moment non-equilibrium coefficient is relaxed independently. How the relaxation factors of higher-order moments affect numerical stability and accuracy remains to be investigated.
\item Except for the recursive regularized models mentioned in item 2, the other regularized models are equivalent to setting the relaxation times of kinetic moments above the highest retained non-equilibrium order to one in raw-moment, central-moment or temperature-scaled central-moment space.
\item Among the regularized models capable of simulating the compressible thermal NSF equations, only the models of Chen et al. \refc{46}, Shan \refc{55} and Li--Shi--Shan \refc{56} perform relaxation in central-moment or temperature-scaled central-moment space and satisfy Galilean invariance of the thermal conductivity.
\end{enumerate}

The underlying principle of a regularized multiple-relaxation collision model is to truncate higher-order kinetic moments that the discrete velocities cannot represent. This prevents the associated spurious modes from contaminating the hydrodynamic equations. Whether non-unit relaxation factors for these higher-order moments improve numerical stability remains unresolved. The advantages of regularized formulations over other collision models also require firmer theoretical justification, particularly for turbulence, aeroacoustics, multiphase flow and rarefied-gas flow. These applications therefore remain important directions for further development.

\paragraph{Acknowledgements.}
This work was supported by the Fundamental Research Funds for the Central Universities at Harbin Engineering University (3072021CFJ0101) and the Joint Guidance Project of the Natural Science Foundation of Heilongjiang Province (LH2021A007).

\appendix
\section{MRT forcing in raw-moment space}\label{sec:appendix-a}
\renewcommand{\theequation}{A\arabic{equation}}
\setcounter{equation}{0}

The following three appendices give implementation forms of the force-inclusive collision operator in the three Hermite spaces. Non-equilibrium contributions to the force are neglected, as in \Es{56}{61}. Overbars are retained here to distinguish the transformed distribution \E{87} from the physical distribution.

In raw-moment space, the moment evolution is
\begin{equation}
\begin{aligned}
\symbf{\abar}_p^{(n)}={}&\symbf{\abar}^{(n)}
-\frac{\symbf{a}^{(n)}-\symbf{a}_0^{(n)}}{\tau_n}
+\symbf{a}_F^{(n)}.
\end{aligned}
\label{eq:A1}
\end{equation}
The transformed moment is
\begin{equation}
\symbf{\abar}^{(n)}=\symbf{a}^{(n)}
+\frac{\symbf{a}^{(n)}-\symbf{a}_0^{(n)}}{2\tau_n}
-\frac12\symbf{a}_F^{(n)}.
\label{eq:A2}
\end{equation}
Eliminating the physical non-equilibrium moment gives
\begin{equation}
\begin{aligned}
\symbf{\abar}_p^{(n)}={}&\symbf{a}_0^{(n)}
+(1-r_n)\symbf{\abar}_1^{(n)}
+\left(1-\frac{r_n}{2}\right)\symbf{a}_F^{(n)},\\
&r_n=\lambda_n^{-1},\qquad\lambda_n=\tau_n+\tfrac12,
\qquad n\geq1.
\end{aligned}
\label{eq:A3}
\end{equation}
Here $\symbf{\abar}_1^{(n)}=\symbf{\abar}^{(n)}-\symbf{a}_0^{(n)}$ and $\symbf{\abar}_p^{(0)}=\rho$. With $\Fv=\rho\gv$, the force coefficients are
\begin{subequations}\label{eq:A4}
\begin{align}
\symbf{a}_F^{(1)}&=\Fv,\label{eq:A4a}\\
\symbf{a}_F^{(2)}&=2\Fv\uv,\label{eq:A4b}\\
\symbf{a}_F^{(3)}&=3\Fv[\uv\uv+(\theta-1)\delv],\label{eq:A4c}\\
\symbf{a}_F^{(4)}&=4\Fv[\uv\uv\uv+3(\theta-1)\uv\delv].\label{eq:A4d}
\end{align}
\end{subequations}

For clarity, define the post-collision non-equilibrium coefficient by $\symbf{A}^{(n)}=\symbf{\abar}_p^{(n)}-\symbf{a}_0^{(n)}$, with $\symbf{A}^{(0)}=0$. It includes the surviving non-equilibrium contribution and the force increment; it is not the coefficient of the physical collision term $\Omega$ alone. Since $\symbf{\abar}_1^{(1)}=-\Fv/2$, its first four orders are
\begin{subequations}
\begin{align}
\symbf{A}^{(1)}&=\frac{\Fv}{2},\label{eq:A5a}\\
\symbf{A}^{(2)}&=(1-r_2)\symbf{\abar}_1^{(2)}
+\left(1-\frac{r_2}{2}\right)\symbf{a}_F^{(2)},\label{eq:A5b}\\
\symbf{A}^{(3)}&=(1-r_3)\symbf{\abar}_1^{(3)}
+\left(1-\frac{r_3}{2}\right)\symbf{a}_F^{(3)},\label{eq:A5c}\\
\symbf{A}^{(4)}&=(1-r_4)\symbf{\abar}_1^{(4)}
+\left(1-\frac{r_4}{2}\right)\symbf{a}_F^{(4)}.\label{eq:A5d}
\end{align}
\end{subequations}

The second- and third-order non-equilibrium coefficients are evaluated by projection. The fourth-order coefficient may also be projected. Alternatively, imposing $\symbf{\dbar}_1^{(4)}=0$ gives the additional regularization assumption
\[
\begin{aligned}
\symbf{\abar}_1^{(4)}={}&4\uv\symbf{\abar}_1^{(3)}
-6\uv_1\symbf{\abar}_1^{(2)}
+4\uv\uv_2\symbf{\abar}_1^{(1)},\\
\uv_1={}&\uv\uv-(\theta-1)\delv,\\
\uv_2={}&\uv\uv-3(\theta-1)\delv.
\end{aligned}
\]
This reconstruction is distinct from the choice of the space in which the moments are relaxed.

Retaining Hermite terms through fourth order, reconstruct the transformed post-collision populations as
\begin{equation}
\bar f_{p,i}=\omega_i\sum_{n=0}^{4}\frac{1}{n!}
\bigl(\symbf{a}_0^{(n)}+\symbf{A}^{(n)}\bigr):\mathcal{H}^{(n)}(\xv_i).
\label{eq:A6}
\end{equation}
The same transformed populations are then streamed:
\begin{equation}
\bar f_i(\boldsymbol{x}+\xv_i,t+1)=\bar f_{p,i}(\boldsymbol{x},t).
\label{eq:A7}
\end{equation}

\section{MRT forcing in central-moment space}\label{sec:Appendix 2}
\renewcommand{\theequation}{B\arabic{equation}}
\setcounter{equation}{0}

Define $\symbf{B}^{(n)}=\symbf{\bbar}_p^{(n)}-\symbf{b}_0^{(n)}$ in the same manner. Then
\begin{equation}
\symbf{B}^{(n)}=(1-r_n)\symbf{\bbar}_1^{(n)}
+\left(1-\frac{r_n}{2}\right)\symbf{b}_F^{(n)}.
\label{eq:B1}
\end{equation}
The force coefficients are $\symbf{b}_F^{(1)}=\Fv$, $\symbf{b}_F^{(2)}=0$, $\symbf{b}_F^{(3)}=3(\theta-1)\Fv\delv$ and $\symbf{b}_F^{(4)}=0$. Retaining only the first-order coefficient is consistent with the equilibrium-force approximation only when $\theta=1$. The third-order force contribution is retained below for general temperature.

Because the first-order non-equilibrium moment of the transformed distribution is nonzero, the transformations are
\begin{subequations}
\begin{align}
\symbf{\bbar}_1^{(1)}&=\symbf{\abar}_1^{(1)},\label{eq:B2a}\\
\symbf{\bbar}_1^{(2)}&=\symbf{\abar}_1^{(2)}-2\uv\symbf{\abar}_1^{(1)},\label{eq:B2b}\\
\symbf{\bbar}_1^{(3)}&=\symbf{\abar}_1^{(3)}-3\uv\symbf{\abar}_1^{(2)}
+3\uv\uv\symbf{\abar}_1^{(1)},\label{eq:B2c}\\
\symbf{\bbar}_1^{(4)}&=\symbf{\abar}_1^{(4)}-4\uv\symbf{\abar}_1^{(3)}\nonumber\\
&\quad+6\uv\uv\symbf{\abar}_1^{(2)}-4\uv\uv\uv\symbf{\abar}_1^{(1)}.\label{eq:B2d}
\end{align}
\end{subequations}
The same transformation applies to $\symbf{A}^{(n)}$ and $\symbf{B}^{(n)}$. Consequently,
\begin{subequations}
\begin{align}
\symbf{A}^{(1)}&=\symbf{B}^{(1)}=\frac{\Fv}{2},\label{eq:B3a}\\
\symbf{A}^{(2)}-2\uv\symbf{A}^{(1)}
&=(1-r_2)\symbf{\bbar}_1^{(2)},\label{eq:B3b}\\
\symbf{A}^{(3)}-3\uv\symbf{A}^{(2)}
&+3\uv\uv\symbf{A}^{(1)}\nonumber\\
&=(1-r_3)\symbf{\bbar}_1^{(3)}
+\left(1-\frac{r_3}{2}\right)\symbf{b}_F^{(3)},\label{eq:B3c}\\
\symbf{A}^{(4)}-4\uv\symbf{A}^{(3)}
&+6\uv\uv\symbf{A}^{(2)}-4\uv\uv\uv\symbf{A}^{(1)}\nonumber\\
&=(1-r_4)\symbf{\bbar}_1^{(4)}.\label{eq:B3d}
\end{align}
\end{subequations}
For an implementation in raw-moment space, introduce the auxiliary coefficients
\[
\symbf{R}^{(n)}=\symbf{\abar}_1^{(n)}+\frac12\symbf{a}_F^{(n)},
\qquad\symbf{R}^{(0)}=\symbf{R}^{(1)}=0.
\]
They combine each transformed non-equilibrium coefficient with its half-force correction and make the relaxation corrections compact. Eliminating the central moments gives
\begin{subequations}
\begin{align}
\symbf{A}^{(1)}&=\frac{\Fv}{2},\label{eq:B4a}\\
\symbf{A}^{(2)}&=(1-r_2)\symbf{\abar}_1^{(2)}
+\left(1-\frac{r_2}{2}\right)\symbf{a}_F^{(2)},\label{eq:B4b}\\
\symbf{A}^{(3)}&=(1-r_3)\symbf{\abar}_1^{(3)}
+\left(1-\frac{r_3}{2}\right)\symbf{a}_F^{(3)}\nonumber\\
&\quad+3(r_3-r_2)\uv\symbf{R}^{(2)},\label{eq:B4c}\\
\symbf{A}^{(4)}&=(1-r_4)\symbf{\abar}_1^{(4)}
+\left(1-\frac{r_4}{2}\right)\symbf{a}_F^{(4)}\nonumber\\
&\quad+4(r_4-r_3)\uv\symbf{R}^{(3)}\nonumber\\
&\quad-6(r_4-2r_3+r_2)\uv\uv\symbf{R}^{(2)}.\label{eq:B4d}
\end{align}
\end{subequations}
Expanding $\symbf{R}^{(n)}$ reproduces the separate non-equilibrium and force corrections. These corrections vanish when all relaxation rates are equal. Setting $r_4=1$ removes the fourth-order central non-equilibrium contribution because $\symbf{b}_F^{(4)}=0$.

The transformed populations are reconstructed and streamed as
\begin{equation}
\bar f_{p,i}=\omega_i\sum_{n=0}^{4}\frac{1}{n!}
\bigl(\symbf{a}_0^{(n)}+\symbf{A}^{(n)}\bigr):\mathcal{H}^{(n)}(\xv_i),
\label{eq:B5}
\end{equation}
\begin{equation}
\bar f_i(\boldsymbol{x}+\xv_i,t+1)=\bar f_{p,i}(\boldsymbol{x},t).
\label{eq:B6}
\end{equation}

\section{MRT forcing in temperature-scaled central-moment space}\label{sec:Appendix 3}
\renewcommand{\theequation}{C\arabic{equation}}
\setcounter{equation}{0}

Let $\symbf{D}^{(n)}=\symbf{\dbar}_p^{(n)}-\symbf{d}_0^{(n)}$. With the equilibrium-force approximation, only $\symbf{d}_F^{(1)}=\Fv/\theta^{(D+1)/2}$ is nonzero. Hence $\symbf{D}^{(1)}=\Fv/[2\theta^{(D+1)/2}]$, while $\symbf{D}^{(n)}=(1-r_n)\symbf{\dbar}_1^{(n)}$ for $n\geq2$. Using $\uv_1$ and $\uv_2$ defined in Appendix~\ref{sec:appendix-a}, the raw-moment form is
\begin{subequations}
\begin{align}
\symbf{A}^{(1)}&=\theta^{(D+1)/2}\symbf{D}^{(1)}
=\frac{\Fv}{2},\label{eq:C1a}\\
\symbf{A}^{(2)}&=\theta^{(D+2)/2}(1-r_2)\symbf{\dbar}_1^{(2)}
+2\uv\symbf{A}^{(1)}\nonumber\\
&=(1-r_2)\bigl(\symbf{\abar}_1^{(2)}-2\uv\symbf{\abar}_1^{(1)}\bigr)
+2\uv\symbf{A}^{(1)},\label{eq:C1b}\\
\symbf{A}^{(3)}&=\theta^{(D+3)/2}(1-r_3)\symbf{\dbar}_1^{(3)}\nonumber\\
&\quad+3\uv\symbf{A}^{(2)}-3\uv_1\symbf{A}^{(1)}\nonumber\\
&=(1-r_3)\bigl(\symbf{\abar}_1^{(3)}-3\uv\symbf{\abar}_1^{(2)}
+3\uv_1\symbf{\abar}_1^{(1)}\bigr)\nonumber\\
&\quad+3\uv\symbf{A}^{(2)}-3\uv_1\symbf{A}^{(1)},\label{eq:C1c}\\
\symbf{A}^{(4)}&=\theta^{(D+4)/2}(1-r_4)\symbf{\dbar}_1^{(4)}\nonumber\\
&\quad+4\uv\symbf{A}^{(3)}-6\uv_1\symbf{A}^{(2)}
+4\uv\uv_2\symbf{A}^{(1)}\nonumber\\
&=(1-r_4)\bigl[\symbf{\abar}_1^{(4)}-4\uv\symbf{\abar}_1^{(3)}\nonumber\\
&\qquad+6\uv_1\symbf{\abar}_1^{(2)}-4\uv\uv_2\symbf{\abar}_1^{(1)}\bigr]\nonumber\\
&\quad+4\uv\symbf{A}^{(3)}-6\uv_1\symbf{A}^{(2)}
+4\uv\uv_2\symbf{A}^{(1)}.\label{eq:C1d}
\end{align}
\end{subequations}

In terms of the auxiliary $\symbf{R}^{(n)}$ from Appendix~\ref{sec:Appendix 2}, the explicit implementation form becomes
\begin{subequations}
\begin{align}
\symbf{A}^{(1)}&=\frac{\Fv}{2},\label{eq:C2a}\\
\symbf{A}^{(2)}&=(1-r_2)\symbf{\abar}_1^{(2)}
+\left(1-\frac{r_2}{2}\right)\symbf{a}_F^{(2)},\label{eq:C2b}\\
\symbf{A}^{(3)}&=(1-r_3)\symbf{\abar}_1^{(3)}
+\left(1-\frac{r_3}{2}\right)\symbf{a}_F^{(3)}\nonumber\\
&\quad+3(r_3-r_2)\uv\symbf{R}^{(2)},\label{eq:C2c}\\
\symbf{A}^{(4)}&=(1-r_4)\symbf{\abar}_1^{(4)}
+\left(1-\frac{r_4}{2}\right)\symbf{a}_F^{(4)}\nonumber\\
&\quad+4(r_4-r_3)\uv\symbf{R}^{(3)}\nonumber\\
&\quad-6(r_4-2r_3+r_2)\uv\uv\symbf{R}^{(2)}\nonumber\\
&\quad+6(r_4-r_2)(\theta-1)\delv\symbf{R}^{(2)}.\label{eq:C2d}
\end{align}
\end{subequations}
The additional fourth-order temperature term distinguishes this result from relaxation in central-moment space. Both reduce to the same expression at $\theta=1$ and to raw-moment BGK relaxation when all rates are equal. Setting $r_4=1$ filters the fourth-order temperature-scaled central non-equilibrium mode.

Finally,
\begin{equation}
\bar f_{p,i}=\omega_i\sum_{n=0}^{4}\frac{1}{n!}
\bigl(\symbf{a}_0^{(n)}+\symbf{A}^{(n)}\bigr):\mathcal{H}^{(n)}(\xv_i),
\label{eq:C3}
\end{equation}
\begin{equation}
\bar f_i(\boldsymbol{x}+\xv_i,t+1)=\bar f_{p,i}(\boldsymbol{x},t).
\label{eq:C4}
\end{equation}

\setlength{\bibsep}{3pt plus 1pt minus 1pt}


\begin{thebibliography}{79}
\small
\setlength{\parskip}{0pt}
\sloppy

\bibitem{r1} CHEN S Y, DOOLEN G D. Lattice Boltzmann method for fluid flows. \textit{Annual Review of Fluid Mechanics}, 1998, 30(1): 329--364. doi: \url{https://doi.org/10.1146/annurev.fluid.30.1.329}.

\bibitem{r2} AIDUN C K, CLAUSEN J R. Lattice-Boltzmann method for complex flows. \textit{Annual Review of Fluid Mechanics}, 2010, 42(1): 439--472. doi: \url{https://doi.org/10.1146/annurev-fluid-121108-145519}.

\bibitem{r3} CHAPMAN S, COWLING T G. The mathematical theory of non-uniform gases: An account of the kinetic theory of viscosity, thermal conduction and diffusion in gases. Cambridge: Cambridge University Press, 1970.

\bibitem{r4} LI F, LI Z H, XU J X, et al. Research on adaptation of CFD software based on many-core architecture of 100P domestic supercomputing system. \textit{Computer Science}, 2020, 47(1): 24--30 (in Chinese). doi: \url{https://doi.org/10.11896/jsjkx.181102176}.

\bibitem{r5} WANG X, AOKI T. Multi-GPU performance of incompressible flow computation by lattice Boltzmann method on GPU cluster. \textit{Parallel Computing}, 2011, 37(9): 521--535. doi: \url{https://doi.org/10.1016/j.parco.2011.02.007}.

\bibitem{r6} SHAN X W, CHEN H D. Lattice Boltzmann model for simulating flows with multiple phases and components. \textit{Physical Review E}, 1993, 47(3): 1815--1819. doi: \url{https://doi.org/10.1103/physreve.47.1815}.

\bibitem{r7} SHAN X W, CHEN H D. Simulation of nonideal gases and liquid-gas phase transitions by the lattice Boltzmann equation. \textit{Physical Review E}, 1994, 49(4): 2941--2948. doi: \url{https://doi.org/10.1103/physreve.49.2941}.

\bibitem{r8} SAGAUT P. Toward advanced subgrid models for Lattice-Boltzmann-based large-eddy simulation: Theoretical formulations. \textit{Computers \& Mathematics with Applications}, 2010, 59(7): 2194--2199. doi: \url{https://doi.org/10.1016/j.camwa.2009.08.051}.

\bibitem{r9} MALASPINAS O, SAGAUT P. Consistent subgrid scale modelling for lattice Boltzmann methods. \textit{Journal of Fluid Mechanics}, 2012, 700: 514--542. doi: \url{https://doi.org/10.1017/jfm.2012.155}.

\bibitem{r10} LIN C D, XU A G, ZHANG G C, et al. Double-distribution-function discrete Boltzmann model for combustion. \textit{Combustion and Flame}, 2016, 164: 137--151. doi: \url{https://doi.org/10.1016/j.combustflame.2015.11.010}.

\bibitem{r11} LI X H, JIANG F, HU C H. Analysis of the accuracy and pressure oscillation of the lattice Boltzmann method for fluid-solid interactions. \textit{Computers \& Fluids}, 2016, 129: 33--52. doi: \url{https://doi.org/10.1016/j.compfluid.2016.01.015}.

\bibitem{r12} LI Z, CAO W J, LE TOUZÉ D. On the coupling of a direct-forcing immersed boundary method and the regularized lattice Boltzmann method for fluid-structure interaction. \textit{Computers \& Fluids}, 2019, 190: 470--484. doi: \url{https://doi.org/10.1016/j.compfluid.2019.06.030}.

\bibitem{r13} QIAN Y H, D'HUMIÈRES D, LALLEMAND P. Lattice BGK models for Navier-Stokes equation. \textit{Europhysics Letters (EPL)}, 1992, 17(6): 479--484. doi: \url{https://doi.org/10.1209/0295-5075/17/6/001}.

\bibitem{r14} BHATNAGAR P L, GROSS E P, KROOK M. A model for collision processes in gases. I. Small amplitude processes in charged and neutral one-component systems. \textit{Physical Review}, 1954, 94(3): 511--525. doi: \url{https://doi.org/10.1103/physrev.94.511}.

\bibitem{r15} D'HUMIÈRES D. Generalized lattice-Boltzmann equations. In: Rarefied gas dynamics: Theory and simulations. Washington, DC: AIAA, 1994: 450--458. doi: \url{https://doi.org/10.2514/5.9781600866319.0450.0458}.

\bibitem{r16} D'HUMIÈRES D, GINZBURG I, KRAFCZYK M, et al. Multiple-relaxation-time lattice Boltzmann models in three dimensions. \textit{Philosophical Transactions of the Royal Society of London Series A: Mathematical, Physical and Engineering Sciences}, 2002, 360(1792): 437--451. doi: \url{https://doi.org/10.1098/rsta.2001.0955}.

\bibitem{r17} GEIER M, GREINER A, KORVINK J G. Cascaded digital lattice Boltzmann automata for high Reynolds number flow. \textit{Physical Review E}, 2006, 73(6): 066705. doi: \url{https://doi.org/10.1103/physreve.73.066705}.

\bibitem{r18} GEIER M, GREINER A, KORVINK J G. A factorized central moment lattice Boltzmann method. \textit{The European Physical Journal Special Topics}, 2009, 171(1): 55--61. doi: \url{https://doi.org/10.1140/epjst/e2009-01011-1}.

\bibitem{r19} LYCETT-BROWN D J. Lattice Boltzmann methods for high end computing of multiphase flow [PhD thesis]. Southampton: University of Southampton, 2013. \url{https://eprints.soton.ac.uk/361706/1/Thesis.pdf}.

\bibitem{r20} LYCETT-BROWN D J, LUO K H. Multiphase cascaded lattice Boltzmann method. \textit{Computers \& Mathematics with Applications}, 2014, 67(2): 350--362. doi: \url{https://doi.org/10.1016/j.camwa.2013.08.033}.

\bibitem{r21} LYCETT-BROWN D, LUO K H. Cascaded lattice Boltzmann method with improved forcing scheme for large-density-ratio multiphase flow at high Reynolds and Weber numbers. \textit{Physical Review E}, 2016, 94(5-1): 053313. doi: \url{https://doi.org/10.1103/PhysRevE.94.053313}.

\bibitem{r22} DUBOIS F, FEVRIER T, GRAILLE B. Lattice Boltzmann schemes with relative velocities. \textit{Communications in Computational Physics}, 2015, 17(4): 1088--1112. doi: \url{https://doi.org/10.4208/cicp.2014.m394}.

\bibitem{r23} DUBOIS F, FEVRIER T, GRAILLE B. On the stability of a relative velocity lattice Boltzmann scheme for compressible Navier-Stokes equations. \textit{Comptes Rendus Mécanique}, 2015, 343(10-11): 599--610. doi: \url{https://doi.org/10.1016/j.crme.2015.07.010}.

\bibitem{r24} FEI L L, LUO K H. Consistent forcing scheme in the cascaded lattice Boltzmann method. \textit{Physical Review E}, 2017, 96(5): 053307. doi: \url{https://doi.org/10.1103/physreve.96.053307}.

\bibitem{r25} FEI L, LUO K H, LI Q. Three-dimensional cascaded lattice Boltzmann method: Improved implementation and consistent forcing scheme. \textit{Physical Review E}, 2018, 97(5-1): 053309. doi: \url{https://doi.org/10.1103/physreve.97.053309}.

\bibitem{r26} DE ROSIS A, LUO K H. Role of higher-order Hermite polynomials in the central-moments-based lattice Boltzmann framework. \textit{Physical Review E}, 2019, 99: 013301. doi: \url{https://doi.org/10.1103/physreve.99.013301}.

\bibitem{r27} GEIER M, SCHÖNHERR M, PASQUALI A, et al. The cumulant lattice Boltzmann equation in three dimensions: Theory and validation. \textit{Computers \& Mathematics with Applications}, 2015, 70(4): 507--547. doi: \url{https://doi.org/10.1016/j.camwa.2015.05.001}.

\bibitem{r28} SEEGER S, HOFFMANN H. The cumulant method for computational kinetic theory. \textit{Continuum Mechanics and Thermodynamics}, 2000, 12(6): 403--421. doi: \url{https://doi.org/10.1007/s001610050145}.

\bibitem{r29} GEIER M, PASQUALI A, SCHÖNHERR M. Parametrization of the cumulant lattice Boltzmann method for fourth order accurate diffusion part I: Derivation and validation. \textit{Journal of Computational Physics}, 2017, 348: 862--888. doi: \url{https://doi.org/10.1016/j.jcp.2017.05.040}.

\bibitem{r30} GEIER M, PASQUALI A, SCHÖNHERR M. Parametrization of the cumulant lattice Boltzmann method for fourth order accurate diffusion part II: Application to flow around a sphere at drag crisis. \textit{Journal of Computational Physics}, 2017, 348: 889--898. doi: \url{https://doi.org/10.1016/j.jcp.2017.07.004}.

\bibitem{r31} GRAD H. On the kinetic theory of rarefied gases. \textit{Communications on Pure and Applied Mathematics}, 1949, 2(4): 331--407. doi: \url{https://doi.org/10.1002/cpa.3160020403}.

\bibitem{r32} BANARI A, GEHRKE M, JANßEN C F, et al. Numerical simulation of nonlinear interactions in a naturally transitional flat plate boundary layer. \textit{Computers \& Fluids}, 2020, 203: 104502. doi: \url{https://doi.org/10.1016/j.compfluid.2020.104502}.

\bibitem{r33} NISHIMURA S, HAYASHI K, NAKAYE S, et al. Implicit large-eddy simulation of rotating and non-rotating machinery with cumulant lattice Boltzmann method aiming for industrial applications. In: AIAA Aviation 2019 Forum, Dallas, Texas. Reston, Virginia: AIAA, 2019. doi: \url{https://doi.org/10.2514/6.2019-3526}.

\bibitem{r34} SITOMPUL Y P, AOKI T. A filtered cumulant lattice Boltzmann method for violent two-phase flows. \textit{Journal of Computational Physics}, 2019, 390: 93--120. doi: \url{https://doi.org/10.1016/j.jcp.2019.04.019}.

\bibitem{r35} KARLIN I V, GORBAN A N, SUCCI S, et al. Maximum entropy principle for lattice kinetic equations. \textit{Physical Review Letters}, 1998, 81(1): 6--9. doi: \url{https://doi.org/10.1103/physrevlett.81.6}.

\bibitem{r36} CHIKATAMARLA S S, KARLIN I V. Entropy and Galilean invariance of lattice Boltzmann theories. \textit{Physical Review Letters}, 2006, 97(19): 190601. doi: \url{https://doi.org/10.1103/physrevlett.97.190601}.

\bibitem{r37} KARLIN I V, BÖSCH F, CHIKATAMARLA S S. Gibbs' principle for the lattice-kinetic theory of fluid dynamics. \textit{Physical Review E}, 2014, 90(3): 031302. doi: \url{https://doi.org/10.1103/physreve.90.031302}.

\bibitem{r38} ATIF M, KOLLURU P K, THANTANAPALLY C, et al. Essentially entropic lattice Boltzmann model. \textit{Physical Review Letters}, 2017, 119(24): 240602. doi: \url{https://doi.org/10.1103/physrevlett.119.240602}.

\bibitem{r39} LADD A J C. Numerical simulations of particulate suspensions via a discretized Boltzmann equation. Part 1. Theoretical foundation. \textit{Journal of Fluid Mechanics}, 1994, 271: 285--309. doi: \url{https://doi.org/10.1017/S0022112094001771}.

\bibitem{r40} LADD A J C. Numerical simulations of particulate suspensions via a discretized Boltzmann equation. Part II. Numerical results. \textit{Journal of Fluid Mechanics}, 1994, 271: 311--339. doi: \url{https://doi.org/10.1017/S0022112094001783}.

\bibitem{r41} LADD A J C, VERBERG R. Lattice-Boltzmann simulations of particle-fluid suspensions. \textit{Journal of Statistical Physics}, 2001, 104(5): 1191--1251. doi: \url{https://doi.org/10.1023/A:1010414013942}.

\bibitem{r42} LATT J, CHOPARD B. Lattice Boltzmann method with regularized pre-collision distribution functions. \textit{Mathematics and Computers in Simulation}, 2006, 72(2-6): 165--168. doi: \url{https://doi.org/10.1016/j.matcom.2006.05.017}.

\bibitem{r43} CHEN H, ZHANG R, STAROSELSKY I, et al. Recovery of full rotational invariance in lattice Boltzmann formulations for high Knudsen number flows. \textit{Physica A: Statistical Mechanics and its Applications}, 2006, 362(1): 125--131. doi: \url{https://doi.org/10.1016/j.physa.2005.09.008}.

\bibitem{r44} ZHANG R Y, SHAN X W, CHEN H D. Efficient kinetic method for fluid simulation beyond the Navier-Stokes equation. \textit{Physical Review E}, 2006, 74(4): 046703. doi: \url{https://doi.org/10.1103/physreve.74.046703}.

\bibitem{r45} SHAN X W, YUAN X F, CHEN H D. Kinetic theory representation of hydrodynamics: A way beyond the Navier--Stokes equation. \textit{Journal of Fluid Mechanics}, 2006, 550: 413--441. doi: \url{https://doi.org/10.1017/S0022112005008153}.

\bibitem{r46} CHEN H D, GOPALAKRISHNAN P, ZHANG R Y. Recovery of Galilean invariance in thermal lattice Boltzmann models for arbitrary Prandtl number. \textit{International Journal of Modern Physics C}, 2014, 25(10): 1450046. doi: \url{https://doi.org/10.1142/S0129183114500466}.

\bibitem{r47} SHAN X W, CHEN H D. A general multiple-relaxation-time Boltzmann collision model. \textit{International Journal of Modern Physics C}, 2007, 18(4): 635--643. doi: \url{https://doi.org/10.1142/S0129183107010887}.

\bibitem{r48} MALASPINAS O. Increasing stability and accuracy of the lattice Boltzmann scheme: Recursivity and regularization. arXiv:1505.06900, 2015. \url{https://arxiv.org/pdf/1505.06900.pdf}.

\bibitem{r49} MATTILA K K, PHILIPPI P C, HEGELE L A. High-order regularization in lattice-Boltzmann equations. \textit{Physics of Fluids}, 2017, 29(4): 046103. doi: \url{https://doi.org/10.1063/1.4981227}.

\bibitem{r50} COREIXAS C, WISSOCQ G, PUIGT G, et al. Recursive regularization step for high-order lattice Boltzmann methods. \textit{Physical Review E}, 2017, 96(3): 033306. doi: \url{https://doi.org/10.1103/physreve.96.033306}.

\bibitem{r51} JACOB J, MALASPINAS O, SAGAUT P. A new hybrid recursive regularised Bhatnagar--Gross--Krook collision model for Lattice Boltzmann method-based large eddy simulation. \textit{Journal of Turbulence}, 2018, 19(11-12): 1051--1076. doi: \url{https://doi.org/10.1080/14685248.2018.1540879}.

\bibitem{r52} FENG Y L, BOIVIN P, JACOB J, et al. Hybrid recursive regularized thermal lattice Boltzmann model for high subsonic compressible flows. \textit{Journal of Computational Physics}, 2019, 394: 82--99. doi: \url{https://doi.org/10.1016/j.jcp.2019.05.031}.

\bibitem{r53} GUO S, FENG Y, SAGAUT P. Improved standard thermal lattice Boltzmann model with hybrid recursive regularization for compressible laminar and turbulent flows. \textit{Physics of Fluids}, 2020, 32(12): 126108. doi: \url{https://doi.org/10.1063/5.0033364}.

\bibitem{r54} RENARD F, FENG Y L, BOUSSUGE J F, et al. Improved compressible hybrid lattice Boltzmann method on standard lattice for subsonic and supersonic flows. \textit{Computers \& Fluids}, 2021, 219: 104867. doi: \url{https://doi.org/10.1016/j.compfluid.2021.104867}.

\bibitem{r55} SHAN X W. Central-moment-based Galilean-invariant multiple-relaxation-time collision model. \textit{Physical Review E}, 2019, 100(4): 043308. doi: \url{https://doi.org/10.1103/physreve.100.043308}.

\bibitem{r56} LI X H, SHI Y Y, SHAN X W. Temperature-scaled collision process for the high-order lattice Boltzmann model. \textit{Physical Review E}, 2019, 100: 013301. doi: \url{https://doi.org/10.1103/physreve.100.013301}.

\bibitem{r57} NIU X D, HYODO S A, MUNEKATA T, et al. Kinetic lattice Boltzmann method for microscale gas flows: Issues on boundary condition, relaxation time, and regularization. \textit{Physical Review E}, 2007, 76(3 Pt 2): 036711. doi: \url{https://doi.org/10.1103/PhysRevE.76.036711}.

\bibitem{r58} LATT J, CHOPARD B, MALASPINAS O, et al. Straight velocity boundaries in the lattice Boltzmann method. \textit{Physical Review E}, 2008, 77(5): 056703. doi: \url{https://doi.org/10.1103/physreve.77.056703}.

\bibitem{r59} MALASPINAS O, CHOPARD B, LATT J. General regularized boundary condition for multi-speed lattice Boltzmann models. \textit{Computers \& Fluids}, 2011, 49(1): 29--35. doi: \url{https://doi.org/10.1016/j.compfluid.2011.04.010}.

\bibitem{r60} WISSOCQ G, GOURDAIN N, MALASPINAS O, et al. Regularized characteristic boundary conditions for the Lattice-Boltzmann methods at high Reynolds number flows. \textit{Journal of Computational Physics}, 2017, 331: 1--18. doi: \url{https://doi.org/10.1016/j.jcp.2016.11.037}.

\bibitem{r61} FENG Y, GUO S, JACOB J, et al. Solid wall and open boundary conditions in hybrid recursive regularized lattice Boltzmann method for compressible flows. \textit{Physics of Fluids}, 2019, 31(12): 126103. doi: \url{https://doi.org/10.1063/1.5129138}.

\bibitem{r62} OTOMO H, CROUSE B, DRESSLER M, et al. Multi-component lattice Boltzmann models for accurate simulation of flows with wide viscosity variation. \textit{Computers \& Fluids}, 2018, 172: 674--682. doi: \url{https://doi.org/10.1016/j.compfluid.2018.02.001}.

\bibitem{r63} OTOMO H, ZHANG R Y, CHEN H D. Improved phase-field-based lattice Boltzmann models with a filtered collision operator. \textit{International Journal of Modern Physics C}, 2019, 30(10): 1941009. doi: \url{https://doi.org/10.1142/S0129183119410092}.

\bibitem{r64} BA Y, WANG N N, LIU H H, et al. Regularized lattice Boltzmann model for immiscible two-phase flows with power-law rheology. \textit{Physical Review E}, 2018, 97(3): 033307. doi: \url{https://doi.org/10.1103/physreve.97.033307}.

\bibitem{r65} ZHUO C, SAGAUT P. Acoustic multipole sources for the regularized lattice Boltzmann method: Comparison with multiple-relaxation-time models in the inviscid limit. \textit{Physical Review E}, 2017, 95(6-1): 063301. doi: \url{https://doi.org/10.1103/physreve.95.063301}.

\bibitem{r66} LEW P T, GOPALAKRISHNAN P, CASALINO D, et al. An extended lattice Boltzmann methodology for high subsonic jet noise prediction. In: Proceedings of the 20th AIAA/CEAS Aeroacoustics Conference, Atlanta, GA. Reston, Virginia: AIAA, 2014. doi: \url{https://doi.org/10.2514/6.2014-2755}.

\bibitem{r67} AVALLONE F, VAN DER VELDEN W C P, RAGNI D, et al. Noise reduction mechanisms of sawtooth and combed-sawtooth trailing-edge serrations. \textit{Journal of Fluid Mechanics}, 2018, 848: 560--591. doi: \url{https://doi.org/10.1017/jfm.2018.377}.

\bibitem{r68} ROMANI G, CASALINO D. Rotorcraft blade-vortex interaction noise prediction using the Lattice-Boltzmann method. \textit{Aerospace Science and Technology}, 2019, 88: 147--157. doi: \url{https://doi.org/10.1016/j.ast.2019.03.029}.

\bibitem{r69} GENDRE F. Développement de méthodes de Boltzmann sur réseau en maillages non-uniformes pour l'aéroacoustique automobile [PhD thesis]. Marseille: Aix-Marseille Université, 2018.

\bibitem{r70} ASTOUL T. Towards improved lattice Boltzmann aeroacoustic simulations with non-uniform grids: Application to landing gears noise prediction [PhD thesis]. Marseille: Aix-Marseille Université, 2021.

\bibitem{r71} SHAN X W. The mathematical structure of the lattices of the lattice Boltzmann method. \textit{Journal of Computational Science}, 2016, 17: 475--481. doi: \url{https://doi.org/10.1016/j.jocs.2016.03.002}.

\bibitem{r72} SHAN X W, LI X H, SHI Y Y. A multiple-relaxation-time collision model by Hermite expansion. \textit{Philosophical Transactions of the Royal Society A: Mathematical, Physical and Engineering Sciences}, 2021, 379(2208): 20200406. doi: \url{https://doi.org/10.1098/rsta.2020.0406}.

\bibitem{r73} MALASPINAS O P. Lattice Boltzmann method for the simulation of viscoelastic fluid flows [PhD thesis]. Lausanne, Switzerland: École Polytechnique Fédérale de Lausanne, 2009.

\bibitem{r74} MARTYS N S, SHAN X W, CHEN H D. Evaluation of the external force term in the discrete Boltzmann equation. \textit{Physical Review E}, 1998, 58(5): 6855--6857. doi: \url{https://doi.org/10.1103/physreve.58.6855}.

\bibitem{r75} HE X Y, SHAN X W, DOOLEN G D. Discrete Boltzmann equation model for nonideal gases. \textit{Physical Review E}, 1998, 57(1): R13--R16. doi: \url{https://doi.org/10.1103/physreve.57.R13}.

\bibitem{r76} GUO Z L, ZHENG C G, SHI B C. Discrete lattice effects on the forcing term in the lattice Boltzmann method. \textit{Physical Review E}, 2002, 65(4): 046308. doi: \url{https://doi.org/10.1103/physreve.65.046308}.

\bibitem{r77} LI X H, SHAN X W. Rotational symmetry of the multiple-relaxation-time collision model. \textit{Physical Review E}, 2021, 103(4): 043309. doi: \url{https://doi.org/10.1103/physreve.103.043309}.

\bibitem{r78} DELLAR P J. Incompressible limits of lattice Boltzmann equations using multiple relaxation times. \textit{Journal of Computational Physics}, 2003, 190(2): 351--370. doi: \url{https://doi.org/10.1016/S0021-9991(03)00279-1}.

\bibitem{r79} CHEN H D, ZHANG R Y, GOPALAKRISHNAN P. Filtered lattice Boltzmann collision formulation enforcing isotropy and Galilean invariance. \textit{Physica Scripta}, 2020, 95(3): 034003. doi: \url{https://doi.org/10.1088/1402-4896/ab4b4d}.

\end{thebibliography}
\end{document}